\documentclass{pspum-l}

\usepackage{amssymb}

\usepackage{latexsym,amsmath}
\usepackage{ccaption,url,a4wide}
\usepackage{pdfsync}
\usepackage[numbers,sort&compress]{natbib}
\newtheorem{theorem}{Theorem}[section]

\theoremstyle{definition}
\newtheorem{definition}[theorem]{Definition}

\theoremstyle{remark}

\newtheorem{prop}[theorem]{Proposition}

\numberwithin{equation}{section}
\newcommand{\be}{\begin{equation}}
\newcommand{\ee}{\end{equation}}
\newcommand{\ba}{\begin{eqnarray}}
\newcommand{\ea}{\end{eqnarray}}
\newcommand{\hf}{\frac{1}{2}}
 
\newcommand\req[1]{(\ref{#1})}
\newcommand{\h}{\hat}
\newcommand\F{\mathbb{F}}
\newcommand\C{\mathbb{C}}
\renewcommand\H{\mathbb{H}}

\newcommand\Q{\mathbb{Q}}
\newcommand\R{\mathbb{R}}
\newcommand\Z{\mathbb{Z}}

\newcommand\CCC{\mathcal{C}}
\newcommand\DDD{\mathcal{D}}

\newcommand\III{\mathcal{I}}

\newcommand\OOO{\mathcal{O}}

\newcommand\TTT{\mathcal{T}}
\newcommand\XXX{\mathcal{X}}
\newcommand\YYY{\mathcal{Y}}

\newcommand\spann{\mathop{\mathrm{span}}\nolimits}

\def\id{\mathchoice{\setlength{\unitlength}{1ex}\identitaet}{
\setlength{\unitlength}{1ex}\identitaet}{
\setlength{\unitlength}{0.5ex}
\begin{picture}(1.6,1.5)
\put(0.7,-0.1){\scriptsize 1}\thinlines
\put(1,0.1){\line(0,1){1.4}}
\put(0.2,0){\line(1,0){1.2}}
\put(0.6,1.5){\line(1,0){0.4}}
\end{picture}}{\setlength{\unitlength}{0.5ex}
\begin{picture}(1.6,1.5)
\put(0.7,-0.1){\scriptsize 1}\thinlines
\put(1,0.1){\line(0,1){1.4}}
\put(0.2,0){\line(1,0){1.2}}
\put(0.6,1.5){\line(1,0){0.4}}
\end{picture}}
}
\newcommand\identitaet{\begin{picture}(1.6,1.5)
\put(0,0){1}
\put(1,0.1){\line(0,1){1.4}}
\thinlines
\put(0.2,0){\line(1,0){1.2}}
\put(0.6,1.5){\line(1,0){0.4}}
\end{picture}}

\newcommand\fa{\forall\,}

\newcommand\qu{\overline}

\newcommand\wt{\widetilde}
\newcommand\wh{\widehat}

\captionnamefont{\bfseries}

\begin{document}


\title{
Symmetry-Surfing the Moduli Space of Kummer K3s}


%    author one information
\author{Anne Taormina}
\address{Centre for Particle Theory\\ Department of Mathematical Sciences\\ Durham University\\Durham, DH1 3LE \\U.K. }
\email{anne.taormina@durham.ac.uk}
\thanks{A.T. thanks the University of Freiburg for their hospitality, and acknowledges a Leverhulme Research Fellowship RF/2012-335.}
%%%%%%%%%%%%%%%%%%%%%%%%%%%%%%%%%%%%%%%%%%%%%%%%%%%%%%%%%%%%%%%

%    author two information
\author{Katrin Wendland}
\address{Mathematics Institute\\ University of Freiburg\\ D-79104 Freiburg\\Germany. }
\email{katrin.wendland@math.uni-freiburg.de}
\thanks{K.W. acknowledges an ERC Starting Independent Researcher Grant StG No. 204757-TQFT}
\thanks{We thank Ron Donagi, Matthias Gaberdiel and Roberto Volpato for very helpful discussions.
We also thank the Heilbronn Institute and the International Centre for Mathematical Sciences in Edinburgh 
as well as the (other) organisers of the Heilbronn Day and Workshop on 
`Algebraic geometry, modular forms and applications to physics', where part of this work was done. }

%    The 2010 edition of the Mathematics Subject Classification is
%    now available.  If you are citing a classification from the
%    new scheme, use the following input coding instead.
%\subjclass[2010]{Primary }

\date{September 30, 2013}
\dedicatory{Dedicated to Prof. Dr. Friedrich Hirzebruch, October 17, 1927 - May 27, 2012,\\ 
in admiration and gratitude: \\
To an extraordinary scientist,
an unforgettable teacher, 
and a model of altruism.}

\begin{abstract}
A maximal subgroup of the Mathieu group $M_{24}$ arises as the combined 
holomorphic symplectic automorphism group of all Kummer surfaces whose K\"ahler class
is induced from the underlying complex torus. As a subgroup of $M_{24}$, this group is
the stabilizer group of an octad in the Golay code. To meaningfully combine the symmetry groups
of distinct Kummer surfaces, we introduce the concepts of Niemeier markings and
overarching maps between pairs of Kummer surfaces. The latter induce a prescription for
symmetry-surfing the moduli space, while the former can be seen as a first step
towards constructing a vertex algebra that governs the elliptic genus of K3
in an $M_{24}$-compatible fashion. We thus argue that a geometric approach 
from K3 to Mathieu Moonshine may bear fruit.

\end{abstract}

\maketitle

\section*{Introduction}

This work is motivated by several mysteries related to the Mathieu Moonshine
phenomenon. Central to this phenomenon is the
elliptic genus of K3, which encodes
topological data on K3 surfaces and at the same time is expected to organise a selection of
states in $N=(4,4)$ superconformal field theories (SCFTs) on K3 into representations of 
the Mathieu group $M_{24}$.
The existence of the relevant representations follows
from Gannon's result \cite{ga12}, which in turn builds on 
the work of Cheng, Gaberdiel-Hohenegger-Volpato and Eguchi-Hikami  \cite{ch10, ghv10a,ghv10b,eghi11}.
The precise construction of those
representations in terms of conformal field theory data, however, has been completely elusive so far,
since the detailed nature of the states governing the elliptic genus has not been pinned down. 
Indeed,
the elliptic genus is a topological invariant generalizing the genera of multiplicative
sequences that were introduced by F.~Hirzebruch \cite{hi66}. It
can be viewed as the regularized index of a $U(1)$-equivariant 
Dirac operator on the loop space  of K3 \cite{akmw87,wi87}. It also arises from the supertrace over the subsector 
of Ramond-Ramond states  of every superconformal field theory on K3, and hence it counts 
states with signs \cite{eoty89,ka05}. That the net contribution should
yield a well-defined representation of any group, let alone of $M_{24}$, is mysterious. However, from
the properties of {twining and twisted-twining genera} it has been argued that one should actually expect
this representation to be realized in terms of
a vertex algebra $\wh\XXX$ \cite{gprv12}. We share that view, although not the recent claim by some 
experts exclusively expecting 
holomorphic vertex algebras in this context,
and casting doubts on whether K3 surfaces bear any key to the Mathieu
Moonshine Mysteries \cite{ga12,gpv13}.\\

In fact, we argue 
that the resolution of certain aspects of 
Mathieu Moonshine  might benefit from deepening our understanding of the implications of 
Mukai's work \cite{mu88}, and from building on the insights offered by Kondo \cite{ko98}. 
Of course, Mukai has proved in \cite{mu88} that every holomorphic 
symplectic symmetry group of a K3 surface is a subgroup of the group $M_{24}$. But he
 also proved that all these symmetry groups are smaller than $M_{24}$ by orders of magnitude.
In fact, all of them are subgroups of $M_{23}$. In \cite{tawe11} we advertised the idea that
presumably, $M_{24}$ could be obtained by combining the holomorphic symplectic
symmetry groups of distinct K3 surfaces at
different points of the moduli space. As a test bed, we proved the existence of an overarching
map $\Theta$ which allows  to combine the holomorphic symplectic symmetry groups of
two special, distinct Kummer surfaces  in terms of their induced actions on the 
Niemeier lattice $N$ of type $A_1^{24}$. We also proved that this combined action on $N$ yields
the largest possible group that can arise by means of such an overarching map. This group is $(\Z_2)^4\rtimes A_7$,
which we therefore called the \textsc{overarching finite symmetry group} of Kummer surfaces.
It contains as proper subgroups all holomorphic symplectic symmetry groups of 
 Kummer surfaces which are
equipped with the dual K\"ahler class  induced from the underlying torus.\\

In this note, in Section\,\ref{sec:quaternions} we briefly recall the Kummer construction and 
gather the information appearing in \cite{tawe11} that is useful for the present work.
In Section\,\ref{sec:overarching}, we introduce the concept of Niemeier markings and 
generalize the ideas summarized above by showing that the technique introduced
for two specific examples of Kummer surfaces in \cite{tawe11}, namely the tetrahedral
and the square Kummer K3, generalizes to other pairs of
Kummer surfaces.  As an application of this technique, Section\,\ref{sec:constructionmaps} 
constructs three overarching maps for three pairs of Kummer surfaces with maximal symmetry. 
Section\,\ref{sec:24A8} shows that for any pair of Kummer K3s, one can find representatives in the smooth
universal cover of the moduli space of hyperk\"ahler structures such that there exists an overarching map analogous
to the one constructed in \cite{tawe11}. Moreover, there always exists a continuous path
between the two representatives of our Kummer surfaces, such that $\Theta$ is compatible with all holomorphic
symplectic symmetries along the path. This is the idea of symmetry-surfing the moduli space, alluded to
in the title of the present paper.

Our surfing procedure allows us to combine the action of all
holomorphic symplectic symmetry groups of Kummer surfaces with induced dual K\"ahler class 
by means of their
induced actions on the lattice $N$. In fact, this action is independent of all choices of
overarching maps. We also prove in Section\,\ref{sec:24A8} that the combined action of 
all these groups is given by a faithful representation of
$(\Z_2)^4\rtimes A_8$ on $N$. The subgroup $(\Z_2)^4\rtimes A_7$, i.e.\ the
overarching finite symmetry group of Kummer surfaces, is the stabilizer subgroup
of $(\Z_2)^4\rtimes A_8$
for one root in the Niemeier lattice $N$, just as the subgroup $M_{23}$ of the Mathieu group
$M_{24}$ is the stabilizer subgroup of $M_{24}$, which naturally acts on $N$,
for one root in $N$. We view this as  evidence
that the Mathieu Moonshine phenomenon is tied to the largest Mathieu group $M_{24}$
rather than $M_{23}$, as  also argued by Gannon \cite{ga12}.

In Section\,\ref{sec:interpretation}, we highlight the relevance of our geometric approach, 
and in particular of the Niemeier markings, in the quest for a vertex algebra that governs the elliptic genus of K3
at lowest order. To this effect, we establish a link between our work on Kummer surfaces and a 
special class of $N=(4,4)$ SCFTs at central charge $c=\qu c=6$, namely $\Z_2$-orbifolds of toroidal conformal field 
theories\footnote{To avoid clumsy terminology,
we simply refer to those SCFTs $\CCC$ on K3 which are obtained by the standard $\Z_2$-orbifold
procedure from a toroidal theory  as ``$\Z_2$-orbifolds''.}. 
This necessitates a transition from geometry to superconformal theory language, which we describe
in Appendix\,\ref{sec:transcft}. The upshot is that our surfing idea is natural: the symmetry groups act 
on the twisted ground states of the $\Z_2$-orbifold conformal field theories, and that action completely determines 
these symmetries.
The twisted ground states can be viewed as a stable part of
the Hilbert space when one surfs between $\Z_2$-orbifolds. As such the twisted ground
states collect the various symmetry groups just like the Niemeier lattice does  by means
of our Niemeier markings. In passing we explain how the very idea of constructing a vertex 
algebra from the field content of SCFTs on K3, which simultaneously governs the elliptic genus
and symmetries,  motivates why we restrict our attention to symmetry groups
that are induced from geometric symmetries in some geometric interpretation, that is,
to subgroups of $M_{24}$. 
%
%%%%%%%%%%%%%%%%%%%%%%%%%%%%%%%%%%%%%%%%%%%%%%%%%%%
\section{Kummer surfaces and quaternions}
\label{sec:quaternions}
%%%%%%%%%%%%%%%%%%%%%%%%%%%%%%%%%%%%%%%%%%%%%%%%%%%
An interesting class of K3 surfaces is obtained through the \textsc{Kummer construction}, which amounts to taking 
a $\mathbb{Z}_2$-orbifold of any complex torus $T$ of dimension $2$, and minimally resolving the singularities 
that arise from the orbifold procedure. More specifically, let $T=T(\Lambda)=\mathbb{C}^2/\Lambda$ with 
$\Lambda\subset\C^2$ denote a lattice of 
rank $4$ over $\Z$, and with generators $\vec{\lambda_i},\, i  \in\{1,\ldots,4\}$. The group $\Z_2$ acts 
naturally on $\C^2$ by $(z_1,z_2)\mapsto (-z_1,-z_2)$ and thereby on $T(\Lambda)$. Using Euclidean 
coordinates $\vec{x}=(x_1,x_2,x_3,x_4)$, where $z_1=x_1+ix_2$ and $z_2=x_3+ix_4$, points on the 
quotient $T(\Lambda)/\Z_2$ are identified according to
$$
\vec{x}\sim \vec{x}+\sum_{i=1}^4n_i {\vec\lambda_i},\quad n_i \in \mathbb{Z},\qquad
\vec{x}\sim-\vec{x}.\nonumber
$$
Hence $T(\Lambda)/\Z_2$ has $16$ singularities of type $A_1$, located at the fixed points of 
the $\Z_2$-action. These fixed points are conveniently labelled by the \textsc{hypercube} 
$\mathbb {F}_2^4\cong{1\over2}\Lambda/\Lambda$, where $\mathbb {F}_2=\{0,1\}$ is the finite field with two elements, as
\be\label{labels}
\vec{F}_{\vec a}:=\left[{\textstyle\hf} \sum_{i=1}^4a_i \vec{\lambda_i}\right]\in T(\Lambda)/\Z_2,\quad
\vec{a}=(a_1,a_2,a_3,a_4) \in \mathbb {F}_2^4.
\ee
\begin{definition}\label{defKummer}
The complex surface $X_{\Lambda}$ obtained by minimally resolving the 
$16$ singularities of  $T(\Lambda)/\Z_2$ is a K3 surface 
\mbox{\rm(}see e.g.\ \mbox{\rm\cite{ni75})} called a  \textsc{Kummer surface}\footnote{We denote
by $\pi\colon T\dashrightarrow X$ the corresponding rational map of degree $2$, and by
$\pi_\ast\colon H_\ast(T,\Z)\longrightarrow H_\ast(X,\Z)$ the induced map on homology.}.
\end{definition}
According to the above definition,  the Kummer surface $X_{\Lambda}$ carries the 
complex structure induced from the universal cover $\mathbb{C}^2$ of $T$. It may also be equipped with 
a K\"ahler structure\footnote{For most parts of our work, the K\"ahler class is degenerate in the sense
that it corresponds to an orbifold limit of K\"ahler metrics.}, 
and this is natural if one is interested in the description 
of {\em finite} groups of symplectic automorphisms of Kummer surfaces. 
We specify such a K\"ahler structure by choosing a so-called
\textsc{dual K\"ahler class} $\omega$, that is, a homology class  which is Poincar\'e dual to a K\"ahler class.
Indeed, first recall the following:
\begin{definition}\label{symmgrp}
Consider a K3 surface $X$.
A map $f\colon X\longrightarrow X$ of finite order is called a
\textsc{symplectic automorphism} if and only if $f$ is biholomorphic
and it induces the identity map on $H^{2,0}(X,\C)$.

If $\omega$ is a dual K\"ahler class on $X$ and the induced map 
$f_\ast\colon H_\ast(X,\R)\longrightarrow H_\ast(X,\R)$ leaves $\omega$
invariant, then $f$ is a \textsc{holomorphic symplectic automorphism}
with respect to $\omega$. 

When a dual K\"ahler class $\omega$ on $X$ has been specified, then the group
of holomorphic symplectic {automorphisms} of $X$ with respect to $\omega$ is called
the \textsc{symmetry group} of $X$.
\end{definition}
As an application of the Torelli theorem for K3 surfaces,
the discussion of holomorphic 
symplectic automorphisms $f$ of a K3 surface  $X$ can be entirely rephrased in terms of 
the induced lattice automorphisms $f_\ast$ of the full integral homology lattice 
$H_\ast(X, \mathbb{Z})$ (these and other results on geometry and symmetries of Kummer K3s are standard;
for a summary, see e.g. \cite[Thm.~3.2.2]{tawe11}). 
Then (see \cite[Prop. 3.2.4]{tawe11} for a proof),
\begin{prop}\label{mukaifinite}
Consider a K3 surface $X$, and denote by $G$ a  group of symplectic 
automorphisms of $X$. 
Then $G$ is finite if and only if $X$ possesses a dual K\"ahler class which is invariant under $G$.
\end{prop}

Throughout this work, we focus on Kummer surfaces  $X_{\Lambda,\omega_0}$, 
by which we mean that as K\"ahler structure on $X_{\Lambda}$
we choose the one induced from the standard K\"ahler structure of
the torus $T(\Lambda)$ inherited from  the Euclidean metric on its
universal cover $\C^2$. Here, ${\omega_0}$ denotes the corresponding dual K\"ahler class on $X_\Lambda$. 
This restricts  the symmetry groups of Kummer surfaces that can 
be obtained, but is sufficient to argue for the existence of a combined  symmetry group
$(\mathbb{Z}_2)^4\rtimes A_8$ in Section\,\ref{sec:24A8}.\\

The generic structure of the symmetry group $G$ of the Kummer surface  $X_{\Lambda,{\omega_0}}$ is 
a semi-direct product         
$G=G_t\rtimes G_T$ (see, for example, \cite[Prop.~3.3.4]{tawe11}). 
The normal subgroup $G_t \cong (\mathbb Z_2)^4$ of G is the 
so-called translational automorphism group which is induced from the shifts by half lattice vectors
${1\over2}\vec\lambda,\,\vec\lambda\in\Lambda$,
on the underlying torus $T=T(\Lambda)$.
 The group $G_T$ is the normalizer of $G_t$ in $G$. It is the group of
symmetries of the Kummer surface induced by the holomorphic symplectic automorphisms of the torus $T$ 
fixing $0\in\mathbb{C}^2/{\Lambda}=T$. That is, $G_T\cong G'_T/\mathbb{Z}_2$,
where $G^\prime_T$ is the group of linear holomorphic symplectic automorphisms of $T$. 
These groups and their possible actions on a torus $T$ have been classified by Fujiki
\cite{fu88}, who proves that $G^\prime_T$ is isomorphic to a subgroup of one of the following groups: 
the cyclic groups $\mathbb{Z}_4, \mathbb{Z}_6$, the binary dihedral groups ${\mathcal O}$ and ${\mathcal D}$ of order 
$8$ and $12$, and the binary tetrahedral group ${\TTT}$. This actually implies that the symmetry group 
$G$ is a subgroup of $(\mathbb{Z}_2)^4\rtimes A_6$, where $A_6$ is the alternating group on
six elements.
Moreover\footnote{See the end of this section, items 1.-3., for the precise definitions of the relevant lattices
and group actions.},
$\TTT$ acts only on the so-called tetrahedral torus, while $\mathcal D$ acts only on the so-called triangular torus. 
$\OOO$ can act on the square torus or on the tetrahedral torus, where it is realized as a subgroup of
$\TTT$. Finally the action of the cyclic groups $\mathbb{Z}_4$ and
$\mathbb{Z}_6$ agrees with that of a cyclic subgroup of $\OOO$, ${\mathcal D}$ or  $\TTT$, 
possibly on a torus that does not enjoy the full dihedral or tetrahedral symmetry. In summary,
the maximal groups that can occur are $\OOO$, ${\mathcal D}$ and $\TTT$.

By definition, any element of $G$ must leave the complex structure and the dual K\"ahler class 
${\omega_0}$ of the Kummer surface $X_{\Lambda,\omega_0}$ invariant. Hence 
in terms of real local coordinates $\vec x=(x_1,x_2,x_3,x_4)$ as above and with 
respect to standard real coordinate vector fields $\vec{e}_1,\ldots, \vec{e}_4$, 
using the notations of 
\cite[Section 3]{tawe11}, $G$ must preserve each of the following $2$-cycles 
in $H_2(X_{\Lambda,\omega_0},\R)$,
\be \label{invariant}
\Omega_1=e_1\vee e_3-e_2\vee e_4,\quad \Omega_2=e_1\vee e_4+e_2\vee e_3\,\,\,{\rm and}\,\,\,
\omega_0=e_1\vee e_2+e_3\vee e_4.
\ee
Equivalently, every symmetry group $G$ must preserve the hyperk\"ahler structure which is specified by the
nowhere vanishing holomorphic 2-form 
and the K\"ahler class on $X_{\Lambda,\omega_0}$. We can work
with local holomorphic 
coordinates $(z_1,z_2)$ that are induced from the underlying torus. 
The invariant classes hence are given by $dz_1 \wedge dz_2$, and
${1\over2i}(dz_1\wedge d\qu z_1+dz_2\wedge d\qu z_2)$. Moreover, $G_T \cong G^\prime_T/\mathbb{Z}_2$
where $G^\prime_T$ acts linearly.
In other words, $G_T^\prime$ is a finite subgroup of $SU(2)$. 
Once a group $G^\prime_T\subset SU(2)$ preserving the lattice $\Lambda$ has been identified such 
that $\Z_2\subset G_T^\prime$, then 
$G_T \cong G^\prime_T/\mathbb{Z}_2$ acts faithfully on the Kummer surface $X_{\Lambda,\omega_0}$.\\

It is not surprising that quaternions provide an elegant framework to describe the groups 
$G_T\cong G_T'/\mathbb{Z}_2$ we 
are interested in when symmetry-surfing \cite{fu88,br98}. Indeed, we recall a formalism taken from \cite{br98} which
is tailored to recover the maximal groups $G^\prime_T$ classified by Fujiki, i.e.\ $G'_T\cong {\OOO},{\DDD},{\TTT}$. It moreover
 provides a unified description of the lattice $\Lambda$ for each torus
on which one of these groups can act as automorphism group. 
In fact, each lattice $\Lambda$ is given in terms of unit quaternion generators, and
the automorphisms act by quaternionic left multiplication.

The link between the skew field of quaternions $\H$ and lattices $\Lambda \subset \mathbb{R}^4$ 
is through the natural isomorphism
\be\label{quaternionid}
\mathbb{R}^4\longrightarrow \mathbb{H},\quad q=(q_0,q_1,q_2,q_3) \longmapsto q_0+q_1i+q_2j-q_3k,
\ee
with $\mathbb{H}=\{q=q_0+q_1i+q_2j+q_3k\mid q_\mu \in 
\mathbb{R} , \mu \in\{0,\ldots,3\} \}$.
The unit quaternions  form a group which is isomorphic to $SU(2)$, and 
under the identification \req{quaternionid} its
regular representation on $\R^4\cong\C^2$ is realized by left multiplication
on $\H\cong\R^4$. 
One immediately checks that with this faithful representation, every unit quaternion leaves the standard
holomorphic two-form $dz_1 \wedge dz_2$ and K\"ahler class 
${1\over2i}(dz_1\wedge d\qu z_1+dz_2\wedge d\qu z_2)$ on $\R^4\cong\C^2$ invariant. 
Hence this identification allows us to realize each of our groups $G_T^\prime$ in terms of
a finite group of unit quaternions.

Assume now that $\Lambda\subset \R^4\cong\H$ is a lattice 
of rank $4$ which carries the faithful action of an automorphism group
$G_T^\prime\subset SU(2)$, where $G_T^\prime$ is one of the maximal groups $\OOO,\,{\mathcal D},\,\TTT$
from Fujiki's classification. By the properties of these maximal groups, 
we can assume without loss of generality that $G_T^\prime$ has
generators $a,b,c$ that are represented by unit quaternions of the form 
\begin{eqnarray}\label{generators}
\h a&=&\textstyle
\cos(\frac{\pi}{m})-i\sin(\frac{\pi}{r})+j\cos(\frac{\pi}{n}),\nonumber\\
\h b&=&\textstyle j,\\
\h c&=&\textstyle \cos(\frac{\pi}{n})+j\cos(\frac{\pi}{m})+k\sin(\frac{\pi}{r}),
\nonumber 
\end{eqnarray}
with the constraint $\cos^2(\frac{\pi}{m})+\cos^2(\frac{\pi}{n})=\cos^2(\frac{\pi}{r})$,
where the numbers $m,\,n,\,r\in \Z$ determine the group $G_T^\prime$ \cite{co74}. Moreover, 
for the lattice $\Lambda \subset \mathbb{R}^4 \cong \mathbb{H}$ we can choose the unit quaternion generators $1,\,\h a,\,\h b,\, \h c$. Hence in terms of $\R^4$, we let
\begin{eqnarray*}
\vec{\lambda_1}=(1,0,0,0),\qquad 
\vec{\lambda_2}&=&\textstyle\left(\cos(\frac{\pi}{m}), -\sin(\frac{\pi}{r}), \cos(\frac{\pi}{n}), 0\right),\\
\vec{\lambda_3}=(0,0,1,0),\qquad
\vec{\lambda_4}&=&\textstyle\left(\cos(\frac{\pi}{n}), 0, \cos(\frac{\pi}{m}), -\sin(\frac{\pi}{r})\right),\nonumber
\end{eqnarray*}
be the generators of $\Lambda$. \\

We now summarise the  data needed for symmetry-surfing the  moduli space of Kummer surfaces. 
We describe the three maximal symmetry 
groups $G_T\cong G^\prime_T/\mathbb{Z}_2$ of 
Kummer surfaces induced by the holomorphic symplectic automorphisms of some torus 
$T=T(\Lambda)$ fixing $0\in\mathbb{C}^2/{\Lambda}=T$, along with the possible
lattices $\Lambda$:
\begin{enumerate}
\item \textbf{Dihedral group $\mathbf{D_2\cong {\OOO}/\mathbb{Z}_2 \cong \mathbb{Z}_2 \times \mathbb{Z}_2}$}

Take the lattice $\Lambda$ to be $\Lambda_0:={\rm span}_{\mathbb{Z}}\{1, \h a = i, \h b =j, \h c =k\}$, with $\{\h a ,\h b, \h c\}$ 
generating the quaternionic group $G_T^\prime\cong Q_8$ of order 8. It is immediate that 
$Q_8$ is the automorphism group of
$\Lambda_0$, which is the lattice yielding the square Kummer surface $X_0$ in \cite{tawe11}. 
There, an equivalent description of the generators of the
binary dihedral group ${\OOO}$  was given by
\be \label{generatorssquare}
\alpha_1\colon\quad(z_1,z_2)\longmapsto (iz_1,-iz_2),\qquad
\alpha_2\colon\quad(z_1,z_2)\longmapsto (-z_2,z_1),
\ee
both of which are  of order $4$. 

\item \textbf{Alternating group $\mathbf{A_4\cong {\TTT}/\mathbb{Z}_2}$}%<3,3,2>

The lattice $\Lambda$ may be generated by 
$\{1,\, \h a=\cos(\frac{\pi}{3})-i\sin(\frac{5\pi}{4})+j\cos(\frac{\pi}{3})
,\, \h b=j, \, \h c=\cos(\frac{\pi}{3})+j\cos(\frac{\pi}{3})+k\sin(\frac{5\pi}{4})\}$, 
hence the four lattice vectors that generate $\Lambda$ may be chosen as 
$\vec{\lambda_1}=(1,0,0,0),\, \vec{\lambda_2}=(\hf,\frac{1}{\sqrt{2}},\hf, 0),\,
\vec{\lambda_3}=(0,0,1,0)$ and
$\vec{\lambda_4}=(\hf ,0, \hf, \frac{1}{\sqrt{2}})$. 

One shows that the orbit of $\vec\lambda_1$ under the group $G_T^\prime=\TTT$ yields $24$ unit
 lattice vectors. 
 This lattice is isometric to the lattice $\Lambda_1:=\Lambda_{D_4}$
used in \cite{tawe11} to construct
 the tetrahedral Kummer surface $X_1=X_{D_4}$ from the torus $T(\Lambda_{D_4})$. 
We will use this Kummer surface in what follows, 
hence we recall the generators of $\Lambda_{D_4}$:
 \be\textstyle
 \vec{\lambda_1}=(1,0,0,0),\,\, \vec{\lambda_2}=(0,1,0,0),\,\, \vec{\lambda_3}=(0,0,1,0),\,\, \vec{\lambda_4}=\hf (1,1,1,1).
 \ee
Generators of the binary tetrahedral group $\TTT$ may be taken to be
\ba \label{generatorstetra}
\gamma_1\colon\quad(z_1,z_2)&\longmapsto& (iz_1,-iz_2),\nonumber\\
\gamma_2\colon\quad(z_1,z_2)&\longmapsto& (-z_2,z_1),\\
\gamma_3\colon\quad(z_1,z_2)&\longmapsto& \textstyle\frac{i+1}{2}(i(z_1-z_2), -(z_1+z_2)).\nonumber
\ea
These generators satisfy the relations $\gamma_1^4=\gamma_2^4=\id$ and $\gamma_3^3=\id$. 
Note that the minimum number of generators for the group $\TTT$ is $2$, and indeed, one has
$\gamma_2=\gamma_1^2\gamma_3\gamma_1(\gamma_3)^{-1}$.

\item \textbf{Permutation group $\mathbf{S_3\cong {\DDD}/\mathbb{Z}_2}$}

Take the lattice $\Lambda_2$ generated by 
$\{1,\, \h a=-\cos({\frac{\pi}{3}})+i\sin(\frac{\pi}{3}),\, \h b=j,\,
\h c=-j\cos(\frac{\pi}{3})-k\sin(\frac{\pi}{3}) \}$, hence 
the four lattice vectors that generate 
$\Lambda_2$ may be chosen as 
\be\label{triangularlattice}\textstyle
\vec{\lambda_1}=(1,0,0,0),\quad \vec{\lambda_2}=(-\hf, \frac{\sqrt{3}}{2},0,0),\quad 
\vec{\lambda_3}=(0,0,1,0),\quad\vec{\lambda_4}=(0,0,-\hf,\frac{\sqrt{3}}{2}).
\ee
The orbit of $\vec\lambda_1$ under the binary dihedral group $G_T^\prime\cong\mathcal D$ yields $12$ unit vectors
in $\Lambda_2$. The Kummer surface obtained from $T(\Lambda_2)$ is
the triangular Kummer surface $X_2$.
The generators of $\mathcal D$ have order $3$ and $4$, respectively, and they are given by
\begin{eqnarray} \label{generatorstriang}
\beta_1\colon  (z_1,z_2) &\longmapsto& (\zeta z_1,\zeta^{-1}z_2),\nonumber\\
\beta_2\colon (z_1,z_2) &\longmapsto& (-z_2, z_1),
\end{eqnarray}
where $\zeta:=e^{2\pi i/3}$.
\end{enumerate}
%%%%%%%%%%%%%%%%%%%%%%%%%%%%%%%%%%%%%%%%%%%%%%%%
\section{Overarching maps and Niemeier markings}
\label{sec:overarching}
%%%%%%%%%%%%%%%%%%%%%%%%%%%%%%%%%%%%%%%%%%%%%%%%%%%%%%%%%%%%%%
The description of symmetries of K3 surfaces is most efficient in terms of lattices.
To this end, recall that the geometric action of a  symmetry group $G$ of a K3 
surface $X$  is fully captured by its action on the  lattice $L_{G}=(L^{G})^\perp\cap H_\ast(X,\Z)$, where 
$L^G:=H_\ast(X,\mathbb{Z})^G$. 
This follows from  the Torelli theorem (see the discussion of Def.~\ref{symmgrp}) and 
the very definition of $L^{G}$ as the sublattice of $H_\ast(X,\Z)$ on which $G$ acts trivially. 

On the other hand, if $X_{\Lambda,\omega_0}$ is a Kummer surface with its induced dual K\"ahler class, 
then the induced action of $G$ on the
Kummer lattice $\Pi\subset H_\ast(X,\Z)$ bears all information about the action of $G$ 
(see \cite[Prop.~3.3.3]{tawe11}):
\begin{prop} 
Consider a Kummer surface $X_{\Lambda,\omega_0}$ with its induced 
dual K\"ahler class. Let $\Pi\subset H_\ast(X,\Z)$
denote the Kummer lattice, that is, the smallest
primitive sublattice of the integral K3 homology which contains the $16$ classes
$E_{\vec a}$, $\vec a\in\F_2^4$, that are obtained from blowing up the fixed points 
$\vec F_{\vec a}$ of the $\Z_2$-action on the underlying torus \mbox{\rm\req{labels}}. \\
Then every symmetry
of $X$ induces a permutation of the $E_{\vec a}$. This permutation is given by
an affine linear transformation of  the labels $\vec a\in\F_2^4$, which  in turn uniquely determines the symmetry. 
\end{prop}
In the case of Kummer surfaces we thus have two competing lattices $\Pi$ and $L_G$
which conveniently encode the action of the symmetry group $G$ of $X_{\Lambda,\omega_0}$.
In \cite{tawe11} we argue that neither does $L_{G}$  
contain the rank $16$ Kummer lattice, nor does, in general, the Kummer lattice
contain $L_G$. Instead, combining the two, in \cite[Prop.~3.3.6]{tawe11}
we introduce the lattice $M_G$, which is generated by $L_G$ and  $\Pi$ along with
the vector $\upsilon_0-\upsilon$, where $\upsilon_0,\,\upsilon$ are generators
of $H_0(X,\Z)$ and $H_4(X,\Z)$ with\footnote{On $H_\ast(X,\Z)$, we use the standard
quadratic form which is induced by the intersection form.} $\langle\upsilon_0,\upsilon\rangle=1$. We argue
that in the Kummer case we can generalize and improve some extremely useful techniques introduced  
by Kondo \cite{ko98} to this enlarged lattice $M_G$. Indeed, we prove that this lattice allows a primitive
embedding into the Niemeier lattice $N(-1)$ with root lattice $A_1^{24}$ \cite[Thm.~3.3.7]{tawe11}, 
where the decoration $(-1)$ indicates that the roots of $N(-1)$ have length square $-2$. 
This embedding allows us to view the symmetry group $G$ as a group of lattice automorphisms of $N(-1)$:
the action of $G$ on $N(-1)$ is defined such that the embedding 
$\iota_G\colon M_G\hookrightarrow N(-1)$ is $G$-equivariant, and
$G$ acts trivially on the orthogonal complement of $\iota_G(M_G)$ in $N(-1)$. Since
the automorphism group of $N(-1)$, up to reflections in the roots of $N(-1)$, is the Mathieu group
$M_{24}$, this conveniently realizes every symmetry group $G$ of a Kummer K3 as a subgroup
of $M_{24}$. \\

In what follows, we use the notations and conventions of \cite{tawe11} throughout.
In particular, we fix the Kummer lattice $\Pi$ within the abstract lattice $H_\ast(X,\Z)$  as well as
its image  under $\iota_G$ in $N(-1)$ for every Kummer surface, independently
of the parameters of the underlying torus. More precisely, we fix a unique  \textsc{marking}
for all our Kummer surfaces, that is, an explicit isometry of the lattice $H_\ast(X,\Z)$
with a standard even, unimodular lattice of signature $(4,20)$. As is explained in 
\cite[Sect.~2.2]{tawe11}, the Kummer construction induces a natural such marking, which
in particular fixes the position of $\Pi$ within the lattice $H_\ast(X,\Z)$. In this setting, 
among the data specifying each  Kummer surface we have to include the choice of
generators $\vec\lambda_1,\ldots,\vec\lambda_4\in\R^4$ for the lattice $\Lambda$ of the underlying
torus $T=T(\Lambda)$. Note that the choice of such a fixed marking amounts to the transition
to a smooth universal cover of the moduli space of hyperk\"ahler structures on K3. 
Similarly to $\Pi\subset H_\ast(X,\Z)$, we also fix the position of $\wt\Pi(-1):=\iota_G(\Pi)$ in
$N(-1)$ such that $\wt\Pi$ is  common to all Kummer surfaces. To do so, 
in \cite[(2.14)]{tawe11} we construct a bijection
$I: \III\setminus \OOO_9 \longrightarrow \mathbb{F}_2^4$  between the 
$16$ elements of the set ${\III}:=\{1,2,\dots,24\}$ that do not belong to our choice of 
reference octad 
${\OOO}_9:=\{3,5,6,9,15,19,23,24\}$ from the Golay code
and the vertices of the hypercube $\mathbb{F}_2^4$. In \cite[Prop.~2.3.4]{tawe11} we prove 
that the $\Q$-linear extension of $\iota_G(E_{\vec a}):=  f_{I^{-1}(\vec a)}$ yields an isometry between
$\Pi$ and $\wt\Pi(-1)$, where $\left\{f_n, \,n\in\III\right\}$, denotes a root basis of the
root lattice $A_1^{24}$ in $N(-1)$. Thus we have fixed the position of $\wt\Pi$ within $N$ for all Kummer
surfaces, similarly to fixing the position of $\Pi$ within the abstract lattice $H_\ast(X,\Z)$. This
motivates the
\begin{definition}\label{niemeiermarking}
With notations as above, for a Kummer surface $X_{\Lambda,\omega_0}$ with symmetry group $G$,  an
isometric embedding $\iota_G\colon M_G\hookrightarrow N(-1)$ such that 
$\iota_G(E_{\vec a})=  f_{I^{-1}(\vec a)}$ for all $\vec a\in\F_2^4$ is called a
\textsc{Niemeier marking}.
\end{definition}
By the above, every Kummer surface $X$ allows a Niemeier marking \cite[Prop.~4.1.1]{tawe11}.
In general, the embedding $\iota_G$ is not uniquely determined. However,
the action of $G$ on $N$, which is induced by the requirement that 
$\iota_{G}$ is $G$-equivariant,
is independent of all choices: indeed, $\iota_{G}(E_{\vec a}) = f_{I^{-1}(\vec a)}$ 
$\fa\vec a\in\F_2^4$ fixes the action of
$G$ on the lattice $\wt\Pi\subset N$, and by the arguments presented in the discussion of
\cite[Cor. 3.3.8]{tawe11} this already uniquely determines the action of $G$ on all of $N$. 

In particular, consider
the translational symmetry 
group $G_t\cong (\mathbb{Z}_2)^4$ discussed in Section\,\ref{sec:quaternions}. Its action 
on the roots $f_n,\,n\in\III$, of $N$, which 
is common to all Kummer surfaces,
is generated by the following permutations \cite[Prop.~4.1.1]{tawe11}:
\be\label{genericc24}
G_t:=(\mathbb{Z}_2)^4:\,\,\,
\left\{ 
\begin{array}{rcl}
\iota_1&=& (1,11)(2,22)(4,20)(7,12)(8,17)(10,18)(13,21)(14,16),\\[5pt]
\iota_2&=& (1,13)(2,12)(4,14)(7,22)(8,10)(11,21)(16,20)(17,18),\\[5pt]
\iota_3&=& (1,14)(2,17)(4,13)(7,10)(8,22)(11,16)(12,18)(20,21),\\[5pt]
\iota_4&=& (1,17)(2,14)(4,12)(7,20)(8,11)(10,21)(13,18)(16,22).
\end{array}\right.
\ee
Now recall Mukai's seminal result \cite{mu88} that the symmetry group of every
K3 surface is isomorphic to a subgroup of one of eleven subgroups of the Mathieu group 
$M_{24}$, the largest one of which has $960$ elements. Hence symmetry groups of 
K3 surfaces are by orders of magnitude smaller than the group $M_{24}$, whose
appearance one expects from Mathieu Moonshine. Therefore, in \cite{tawe11} we propose
to use Niemeier markings to combine the symmetry groups of distinct
Kummer surfaces by means of their actions on the Niemeier lattice $N$. To underpin
this idea by lattice identifications, we propose to extend a given Niemeier marking $\iota_G$
to a linear bijection $\Theta: H_\ast(X,\mathbb{Z}) \longrightarrow N(-1)$, which restricts 
to an isometry on the largest possible sublattice of $H_\ast(X,\mathbb{Z})$.
More precisely, we propose to construct a map
$\Theta$   which  induces Niemeier markings of
all K3 surfaces along a smooth path in the smooth universal cover of the moduli space
of hyperk\"ahler structures on K3. If this path connects two distinct Kummer K3s $X_A$ and
$X_B$, then we call $\Theta_{AB}$ an \textsc{overarching map} for $X_A$ and
$X_B$. This is the key to exhibit an overarching symmetry in 
the moduli space of 
Kummer K3s. We say that an overaching map $\Theta_{AB}$ for Kummer surfaces $X_A$ and
$X_B$ allows us to \textsc{surf} from one of the corresponding Kummer surfaces to the other 
\textsc{in moduli space}.

For two Kummer surfaces $X_A,\, X_B$ with complex and K\"ahler structures
induced from the underlying torus and with symmetry groups
$G_A,\, G_B$, respectively,
we will argue below that the following holds:
under appropriate additional assumptions, one can construct an
overarching map $\Theta_{AB}$  which restricts to a Niemeier marking, that is to an isometric $G_k$-equivariant
embedding $\iota_{G_k}\colon M_{G_k}\hookrightarrow N(-1)$, for both $k=A$ and $k=B$,
just like the map $\Theta$
constructed in \cite{tawe11} for the tetrahedral Kummer surface $X_1=X_{D_4}$ and the
square Kummer surface $X_0$. That $\Theta$ restricts
to the desired Niemeier markings
is sufficient to ensure that $\Theta_{AB}$ is an overarching map according to the above definition. 
Indeed,
 we can always find a path in the smooth universal cover of the moduli space which connects
$X_A$ and $X_B$, such that all intermediate points of the path are Kummer surfaces with the minimal symmetry
group $G=G_t\cong(\Z_2)^4$. The group $G_t$ is compatible with $\Theta_{AB}$ by construction.
See \cite[Thm.~4.4.2]{tawe11} for an example -- one 
solely needs to ensure that 
$\spann_\C\{\Omega_1,\,\Omega_2,\,\omega_0\}^\perp\cap\pi_\ast H_2(T,\Z)=\{0\}$ along the path.\\

To determine sufficient conditions on the existence of $\Theta_{AB}$, first note that
by the above, see also
\cite[Thm. 3.3.7]{tawe11}, the lattices $M_{G_k}$ share the Kummer lattice
$\Pi$ and the vector $\upsilon_0-\upsilon$. By the Definition \ref{niemeiermarking} of
 Niemeier markings $\iota_G\colon M_G\hookrightarrow N(-1)$, we require $\Theta_{AB}(E_{\vec a})=f_{I^{-1}(\vec a)}$
for all $\vec a\in\F_2^4$. As mentioned above, $G_k$-equivariance of $\iota_{G_k}$ then already fixes the action
of $G_k$ on $N$. An overarching map $\Theta_{AB}$ hence only exists if 
there is an index $n_0\in\OOO_9$, such that $f_{n_0}$ is invariant under the action of both groups
$G_A,\, G_B$, such that  $\Theta_{AB}(\upsilon_0-\upsilon)=f_{n_0}$ is consistent with $G_k$-equivariance.

For the complementary lattices $\wh K_{G_k}:=((\pi_\ast(H_2(T,\Z))^{(G_k)_T})^\perp \cap \pi_\ast H_2(T,\Z)$ 
for $k\in\{A,B\}$ introduced
in \cite[Thm. 3.3.7]{tawe11}, choose bases $I_{i_k, k}^\perp$, $i_k\in\{1,\ldots,N_k\}$, where
$N_k\leq3$ by construction. If all the vectors $I_{1,A}^\perp,\ldots, I_{N_A,A}^\perp,\, 
I_{1,B}^\perp,\ldots, I_{N_B,B}^\perp$ are linearly independent, then we
claim that under one final
assumption we can find an overarching map $\Theta_{AB}$ for $X_A$ and $X_B$ as desired\footnote{We will see below 
that the assumption of linear independence
can be relaxed, but for simplicity of exposition we first consider this case.}. 
Indeed, as in \cite[\S4.1]{tawe11},
for each of the six 
two-cycles\footnote{Recall that 
for $T=T(\Lambda)$, $\lambda_{ij}:=\lambda_i\vee\lambda_j\in H_2(T,\Z)$
denotes the integral two-cycle specified by the lattice vectors $\vec\lambda_i,\,\vec\lambda_j\in\Lambda$.}
$\lambda_{ij}$, we  first choose a set $Q_{ij}\subset\III$ 
of four labels, such that
$$
\Theta_{AB}(\pi_\ast\lambda_{ij}) = \sum_{n\in Q_{ij}} f_n \mod 2N(-1)
$$
is compatible with the required $G_k$-equivariance. In fact, 
for each $\lambda_{ij}$, this constraint only leaves a choice between two complementary 
sets $Q_{ij}\subset\OOO_9$ which are explicitly listed in \cite[(4.3)]{tawe11}.
Choose these quadruplets of labels
such that for each $Q_{ij}$, $n_0\not\in Q_{ij}$. Analogously to 
\cite[Prop. 4.2.5]{tawe11} this defines a map $\qu I$ through $\qu I(\pi_\ast\lambda_{ij}):=Q_{ij}$
and $\qu I(\lambda+\lambda^\prime):=\qu I(\lambda)+\qu I(\lambda^\prime)$ by symmetric
differences of sets. Since isometric embeddings $\iota_{G_k}\colon M_{G_k}\hookrightarrow N(-1)$
exist by \cite[Prop. 4.1.1]{tawe11}, we can now find appropriate candidates 
$\Theta_{AB}( I_{i_k, k}^\perp )\in N(-1)$ such that $\Theta_{AB}$ restricts to an isometry on both lattices 
$\wh K_{G_k}$. Indeed, up to contributions of the form $2\Delta$ with $\Delta\in N(-1)$, each
$\Theta_{AB}( I_{i_k, k}^\perp )$ is a linear combination of roots $f_j$ with $j\in\qu I( I_{i_k, k}^\perp )$. 
Under the final assumption that all the $\Theta_{AB}( I_{i_k, k}^\perp )$ constructed in this manner
are linearly independent,  clearly $\Theta_{AB}$
can be extended to an overarching map as desired. 

All our assumptions hold true in two of the three
 cases for which we shall construct overarching maps and exhibit overarching symmetries in Section\,\ref{sec:constructionmaps}
below. In one case,  the vectors $I_{1,A}^\perp,\ldots, I_{N_A, A}^\perp,\, 
I_{1, B}^\perp,\ldots, I_{N_B, B}^\perp$ fail to be linearly independent. However, the linear dependence
results from a repetition of vectors, $I_{a, A}^\perp=I_{b, B}^\perp$, so by listing every vector
only once, linear independence is achieved, and the argument goes through as above.\\

This technique allows us to find overarching maps between any two Kummer surfaces, as we
shall see in the next two sections. More precisely, for any pair of Kummer surfaces we can find 
representatives $X_A$ and $X_B$ in the smooth universal cover of the moduli space of 
hyperk\"ahler structures, such that an overarching map for $X_A$ and $X_B$ exists. Hence we
can surf between any two points in moduli space.

%
%%%%%%%%%%%%%%%%%%%%%%%%%%%%%%%%%%%%%%%%%%%%%%%%%%%%%%%%%%%%%%%
\section{Construction of overarching maps}
\label{sec:constructionmaps}
%%%%%%%%%%%%%%%%%%%%%%%%%%%%%%%%%%%%%%%%%%%%%%%%%%%%%%%%%%%%%%
In Section\,\ref{sec:quaternions}, we have identified three distinct Kummer surfaces $X_k,\, k \in\{0, 1, 2\}$, 
whose associated tori $T=\mathbb{C}^2/\Lambda$ have maximal symmetry. In order to explore the 
overarching symmetry for 
 the moduli space of Kummer surfaces by
surfing from $X_0$ to $X_1$ and $X_2$, and from $X_1$ to $X_2$, 
we apply the recipe given in Section\,\ref{sec:overarching} to construct three overarching maps 
$\Theta_{k\ell}$, $0\le k < \ell \le 2$, that  yield overarching symmetry groups 
 for  the three pairs of Kummer surfaces $(X_k, X_{\ell})$.
As was explained in Section\,\ref{sec:overarching}, the construction of an overarching 
map requires the 
existence of a root $f_{n_0} \in N(-1)$,  $n_0 \in {\OOO}_9$, that is invariant under the action of $G_k$ and $G_\ell$. 
In the cases of interest to us here, the value of $n_0$ varies from map to map, but
we carefully note down all possible choices, since this will be crucial in the subsequent section. 
We first summarize the construction of the overarching map $\Theta_{01}$ valid for the square and 
tetrahedral Kummer surfaces, which appeared with some additional details in \cite{tawe11}. Then 
we proceed to the construction of the other two maps, $\Theta_{02}$ and $\Theta_{12}$, which are new. 
This exercise paves the way to Section\,\ref{sec:24A8}, where we argue that one can combine various 
overarching groups and obtain an action
of a maximal subgroup $(\mathbb{Z}_2)^4\rtimes A_8$ of $M_{24}$ on 
the Niemeier lattice $N(-1)$, overarching the 
entire Kummer moduli space.

%%%%%%%%%%%%%%%%%
\subsection{Overarching the square and tetrahedral Kummer K3s}
%%%%%%%%%%%%%%%%%
The full symmetry group of the square Kummer surface $X_0$ is the  group 
$G_0:=(\mathbb{Z}_2)^4\rtimes (\mathbb{Z}_2\times \mathbb{Z}_2)$ of order $64$, while that of the 
tetrahedral Kummer surface $X_1:=X_{D_4}$ is the  group $G_1:=(\mathbb{Z}_2)^4\rtimes A_4$ of order 
$192$. By the discussion in the previous section, there
exist Niemeier markings $\iota_{G_k},\, k\in\{0,\,1\}$, which allow the definition of induced actions of the groups
$G_k$ on the Niemeier lattice $N(-1)$, independently of all choices. Indeed, for the respective generators 
listed at the end of Section\,\ref{sec:quaternions}, according to \cite[Sects.~4.2, 4.3]{tawe11} we obtain
\ba
(G_0)_T:= \mathbb{Z}_2\times \mathbb{Z}_2:&\!\!\!\!\!\!\!\!\!&
\left\{\begin{array}{r@{\!\;}c@{\!\;}l}
\alpha_1&=&(4,8)(6,19)(10,20)(11,13)(12,22)(14,17)(16,18)(23,24),\\[5pt]
\alpha_2&=&(2,21)(3,9)(4,8)(10,12)(11,14)(13,17)(20,22)(23,24),
\end{array}\right. \label{alphapermutations}\\[1em]
(G_1)_T:= A_4:&\!\!\!\!\!\!\!\!\!&
\left\{\begin{array}{r@{\!\;}c@{\!\;}l}
\gamma_1&=&(2,8)(7,18)(9,24)(10,22)(11,13)(12,17)(14,20)(15,19),\\[5pt]
\gamma_2&=&(2,18)(7,8)(9,19)(10,17)(11,14)(12,22)(13,20)(15,24),\\[5pt]
\gamma_3&=&(2,12,13)(4,16,21)(7,17,20)(8,22,14)(9,19,24)(10,11,18).
\end{array}\right.\label{gammapermutations}
\ea
The construction of the map $\Theta_{01}$ requires that one root $f_{n_0}$ 
with $n_0\in {\OOO}_9$ is invariant under $G_0$ and $G_1$. 
One checks that indeed $n_0:=5$ is the only label in $\OOO_9$ which is fixed by both groups.

According to \cite[(4.9),(4.21)]{tawe11},
the generators of the rank $3$ lattices $\wh K_{G_1}$ and $\wh K_{G_0}$ are
\be \label{Iperp01}
\begin{array}{rcrrcr}
I_{1,1}^{\perp}&=&\pi_\ast\lambda_{14}+\pi_\ast\lambda_{24}-\pi_\ast\lambda_{23},
&\qquad 
I_{1,0}^{\perp}&=&\pi_\ast\lambda_{14}-\pi_\ast\lambda_{23},
\\[5pt]
I_{2,1}^{\perp}&=&\pi_\ast\lambda_{13}+\pi_\ast\lambda_{24}+\pi_\ast\lambda_{34},
&\qquad 
I_{2,0}^{\perp}&=&\pi_\ast\lambda_{13}+\pi_\ast\lambda_{24},
\\[5pt]
I_{3,1}^{\perp}&=&-\pi_\ast\lambda_{12}+\pi_\ast\lambda_{14}+\pi_\ast\lambda_{34},
&\qquad 
I_{3,0}^{\perp}&=&\pi_\ast\lambda_{34}-\pi_\ast\lambda_{12}.
\end{array}
\ee
From \cite[(4.3)]{tawe11} we read that $n_0=5\not\in Q_{ij}$ implies
\be \label{Qij5}
\begin{array}{rclrclrcl}
Q_{12}&=&\{3, 6, 15, 19\}, 
&\quad Q_{13}&=&\{6, 15, 23, 24\},
&\quad Q_{14}&=&\{3, 9, 15, 24\}, \\[5pt]
Q_{34}&=&\{6, 9, 15, 19\}, 
  &\quad  Q_{24}&=&\{15, 19, 23, 24\}, 
 &\quad  Q_{23}&=&\{3, 9, 15, 23\}. 
\end{array}
\ee
Hence the map $\qu I$ described in Section\,\ref{sec:overarching} is
\be \label{Ibarmap01}
\begin{array}{rclrclrcl}
\qu I( I_{1,1}^{\perp}) &=& \{ 15, 19\}, 
&\qquad \qu I( I_{2,1}^{\perp}) &=& \{ 9, 15\},
&\qquad \qu I( I_{3,1}^{\perp}) &=& \{ 15, 24\},\\[5pt]
\qu I(I_{1,0}^{\perp}) &=& \{ 23,24\},
&\qquad  \qu I(I_{2,0}^{\perp}) &=& \{ 6, 19\},
 &\qquad \qu I( I_{3,0}^{\perp}) &=& \{ 3, 9\}.
\end{array}
\ee
Our choice of images of the generators \eqref{Iperp01} 
under $\Theta_{01}$ must ensure that $\Theta_{01}$ restricts to an isometry on both lattices 
$\wh K_{G_k}$. 
Therefore, note that the quadratic form on $\wh K_{G_1}$ 
with respect to the basis $I_{i,1}^{\perp},\,i\in\{1,2,3\}$, and that on $\wh K_{G_0}$ with respect to the basis 
$I_{i,0}^{\perp},\, i\in\{1,2,3\}$, are
\ba \label{quadform0}
\wh K_{G_1}:   \left( \begin{array}{rrr}
 -4&-2&-2\\-2&-4&-2\\-2&-2&-4
 \end{array}\right),&&\qquad
\wh K_{G_0}:  \left( \begin{array}{rrr}
 -4&0&0\\0&-4&0\\0&0&-4
 \end{array}\right)
\ea
 according to \cite[(4.20)]{tawe11} and \cite[(4.27)]{tawe11}.
 Then
 the following gives linearly independent
 candidates for the $\Theta_{01}( I_{i_k, k}^\perp )\in N(-1)$ as desired:
\begin{equation} \label{theta01}
\Theta_{01}\colon\quad\left\{
\begin{array}{rcrrcrrcr}
I_{1,1}^{\perp}&\longmapsto&   f_{19}-f_{15},
&\quad I_{2,1}^{\perp}&\longmapsto& f_9-f_{15},
&\quad I_{3,1}^{\perp} &\longmapsto& f_{24}-f_{15},\\[5pt]
I_{1,0}^{\perp} &\longmapsto& f_{24}-f_{23},
&\quad I_{2,0}^{\perp}&\longmapsto& f_{19}-f_{6},
&\quad I_{3,0}^{\perp}&\longmapsto& f_{9}-f_{3} .
\end{array}
\right.
\end{equation}
Equivalently,
\begin{equation}\label{theta01b}
\Theta_{01}\colon\quad\left\{
\begin{array}{rcrcl}
 \pi_\ast \lambda_{12}
&\longmapsto&  2q_{12}&=& f_3+f_6-f_{15}-f_{19},\\[5pt]
 \pi_\ast \lambda_{34}
&\longmapsto& 2q_{34}&=&f_6+f_9-f_{15}-f_{19},\\[5pt]
 \pi_\ast \lambda_{13} 
&\longmapsto& 2q_{13}&=&-f_6+f_{15}-f_{23}+f_{24},\\[5pt]
 \pi_\ast \lambda_{24}
&\longmapsto& 2q_{24}&=&-f_{15}+f_{19}+f_{23}-f_{24},\\[5pt]
 \pi_\ast \lambda_{14}
&\longmapsto& 2q_{14}&=&f_{3}-f_{9}-f_{15}+f_{24},\\[5pt]
 \pi_\ast \lambda_{23}
&\longmapsto& 2q_{23}&=&f_{3}-f_{9}-f_{15}+f_{23}.
\end{array}\right.
\end{equation}

On the Kummer lattice $\Pi$, we set $\Theta_{01}(E_{\vec a})=f_{I^{-1}(\vec a)}$, as always.
Finally, a consistent choice for the images of $\upsilon,\, \upsilon_0$ is
$$
\Theta_{01}\colon\left\{
\begin{array}{rcl}
 \upsilon_0 &\longmapsto& \hf \left(f_3+f_5+f_6+f_9-f_{15}-f_{19}-f_{23}-f_{24}\right),\\[5pt]
 \upsilon &\longmapsto&  \hf \left(f_3-f_5+f_6+f_9-f_{15}-f_{19}-f_{23}-f_{24}\right).
\end{array}
\right.
$$
This completes the construction of the map $\Theta_{01}$ which is compatible with the symmetry groups of the 
square ($G_0)$ and tetrahedral $(G_1)$ Kummer surfaces. Viewed as a linear bijection 
$\Theta_{01}: H_\ast(X_0,\mathbb{Z}) \longrightarrow N(-1)$, its restriction 
$\Theta_{01}\vert_{M_{G_0}}$ yields a $G_0$-equivariant and isometric embedding of 
$M_{G_0}$ in $N(-1)$. Viewed instead as a linear bijection
$\Theta_{01}: H_\ast(X_1,\mathbb{Z}) \longrightarrow N(-1)$, its restriction 
$\Theta_{01}\vert_{M_{G_1}}$ yields a $G_1$-equivariant and isometric embedding of $M_{G_1}$ in $N(-1)$.
This property of the overarching map $\Theta_{01}$ gives us ground to argue that there is an overarching symmetry group
for the square and tetrahedral Kummer surfaces, whose action is encoded in the same Niemeier lattice $N(-1)$ 
through the generators \eqref{genericc24} of the translational symmetry group $G_t$ common to all Kummer 
surfaces, in addition to
the generators \eqref{alphapermutations} {\em and} \eqref{gammapermutations}. The group generated 
this way  is a copy of 
$(\mathbb{Z}_2)^4\rtimes A_7 \subset M_{24}$.

%%%%%%%%%%%%%%%%%%%%%%%%%%%%%%%%%%%%%%%
\subsection{Overarching the square and the triangular Kummer K3s}
%%%%%%%%%%%%%%%%%%%%%%%%%%%%%%%%%%%%%%%
The full symmetry group of the triangular  Kummer  surface $X_2$ is the group 
$G_2:=(\mathbb{Z}_2)^4\rtimes S_3$ of order $96$, see \req{triangularlattice} and \req{generatorstriang}. 
Independently of the choice of 
Niemeier marking, the induced 
action of $G_2$ on the Niemeier lattice is generated by
\be\label{betapermutations}
(G_2)_T:= S_3:\,\,\
\left\{\begin{array}{rcl}
\beta_1&=&(2, 17, 14)(4, 7, 8)(10, 16, 12)(11, 13, 21)(18, 20, 22)(5, 24, 23),\\[5pt]
\beta_2&=&(2,21)(3,9)(4,8)(10,12)(11,14)(13,17)(20,22)(23,24).
\end{array}\right.
\ee
The construction of an overarching map $\Theta_{02}$ for $X_0$ and $X_2$ 
requires a root $f_{n_0}$ with $n_0\in\OOO_9$ which is
 invariant under $G_0$ and $G_2$. 
From \req{alphapermutations} and \req{betapermutations} we observe that $\alpha_2=\beta_2$
and that $n_0=15$ is the only label in $\OOO_9$ which is invariant under both
groups.

To calculate the generators of the lattice $\wh K_{G_2}$ following the techniques
explained in \cite{tawe11}, we first need to determine generators of the lattice $(\pi_\ast H_2(T,\Z))^{(G_2)_T}$. 
With the basis $\vec\lambda_1,\ldots,\vec\lambda_4$
for the triangular lattice given in \req{triangularlattice}, we obtain primitive generators of that lattice as
$$
\pi_\ast \lambda_{13}-\pi_\ast \lambda_{24},\quad
\pi_\ast \lambda_{13}+\pi_\ast \lambda_{23}+\pi_\ast \lambda_{14},\quad
\pi_\ast \lambda_{12}+\pi_\ast\lambda_{34},
$$
and hence the orthogonal complement $\wh K_{G_2}$ of
$(\pi_\ast H_2(T,\Z))^{(G_2)_T}$  in $\pi_\ast H_2(T,\Z)$ is generated by the lattice vectors
\be \label{Iperp02}
I_{1,2}^{\perp}:=\pi_\ast\lambda_{12}-\pi_\ast\lambda_{34},\quad
I_{2,2}^{\perp}:=\pi_\ast\lambda_{13}+\pi_\ast\lambda_{23}+\pi_\ast\lambda_{24},\quad
I_{3,2}^{\perp}:=\pi_\ast\lambda_{14}-\pi_\ast\lambda_{23}.
\ee
From \cite[(4.3)]{tawe11} we read that $n_{0}=15\not\in Q_{ij}$ implies
\be \label{Qij15}
\begin{array}{rclrclrcl}
Q_{12}&=&\{5, 9, 23, 24\}, 
&\quad Q_{13}&=&\{3, 5, 9, 19\},
&\quad Q_{14}&=&\{5, 6, 19, 23\},\\[5pt]
Q_{34}&=&\{3, 5, 23, 24\}, 
 &\quad  Q_{24}&=&\{3, 5, 6, 9\}, 
 &\quad  Q_{23}&=&\{5, 6, 19, 24\}. 
\end{array}
\ee
Hence the map $\qu I$ described in Section\,\ref{sec:overarching} is as 
in \eqref{Ibarmap01} for $I_{i,0}^{\perp},\, i\in\{1,2,3\}$, and furthermore,
$$
 \qu I( I_{1,2}^{\perp}) = \{ 3, 9\},\quad
 \qu I( I_{2,2}^{\perp}) = \{ 5, 24\},\quad
\quad \qu I( I_{3,2}^{\perp}) = \{ 23, 24\}.
$$
We now need to choose the images in $N(-1)$ of the generators 
\eqref{Iperp02} under $\Theta_{02}$ such that $\Theta_{02}$ restricts to an isometry on the lattices $\wh K_{G_0}$ and $\wh K_{G_2}$. 
To do so, note that the quadratic form on $\wh K_{G_0}$ 
with respect to the basis $I_{i,0}^{\perp},\, i\in\{1,2,3\}$, and that of $\wh K_{G_2}$ with 
respect to the basis $I_{i,1}^{\perp},\, i\in\{1,2,3\}$, by \req{quadform0} and \req{Iperp02} are
\be \label{quadform2}
\wh K_{G_0}:  \left( \begin{array}{rrr}
 -4&0&0\\0&-4&0\\0&0&-4
 \end{array}\right),
 \qquad
\wh K_{G_2}:   \left( \begin{array}{rrr}
 -4&0&0\\0&-4&2\\0&2&-4
 \end{array}\right).
\ee
Moreover, we have $I_{1,0}^{\perp}=I_{3,2}^{\perp}$ and $I_{3,0}^{\perp}=I_{1,2}^{\perp}$, such that we can find
candidates for $\Theta_{02}( I_{i_k, k}^\perp )\in N$
as desired:
\begin{equation} \label{theta}
\Theta_{02}\colon\quad
\left\{
\begin{array}{rcrrcrrcr}
I_{1,0}^{\perp}&\longmapsto& f_{24}-f_{23},
&\quad I_{2,0}^{\perp} &\longmapsto& f_{6}-f_{19},
&\quad I_{3,0}^{\perp}&\longmapsto& f_{3}-f_{9},\\[5pt]
I_{1,2}^{\perp}&\longmapsto&   f_{3}-f_{9},
&\quad I_{2,2}^{\perp}&\longmapsto& f_5-f_{24},
&\quad I_{3,2}^{\perp}&\longmapsto& f_{24}-f_{23}.
\end{array}
\right.
\end{equation}

For example, we can choose the following map in order to induce \req{theta}:
\begin{equation}\label{theta02}
\Theta_{02}\colon\quad\left\{
\begin{array}{rcrcl}
 \pi_\ast \lambda_{12}
&\longmapsto&  2q_{12}&=& f_5+f_9-f_{23}-f_{24},\\[5pt]
 \pi_\ast \lambda_{34}
&\longmapsto& 2q_{34}&=&f_3+f_5-f_{23}-f_{24},\\[5pt]
 \pi_\ast \lambda_{13} 
&\longmapsto& 2q_{13}&=&f_3+f_{5}-f_{9}-f_{19},\\[5pt]
\pi_\ast \lambda_{24}
&\longmapsto& 2q_{24}&=&-f_{3}-f_{5}+f_{6}+f_{9},\\[5pt]
 \pi_\ast \lambda_{14}
&\longmapsto& 2q_{14}&=&f_{5}-f_{6}+f_{19}-f_{23},\\[5pt]
 \pi_\ast \lambda_{23}
&\longmapsto& 2q_{23}&=&f_{5}-f_{6}+f_{19}-f_{24}.
\end{array}\right.
\end{equation}
On the Kummer lattice $\Pi$, we set $\Theta_{02}(E_{\vec a})=f_{I^{-1}(\vec a)}$, as before.
Finally, a consistent choice for the images of $\upsilon,\, \upsilon_0$ is
\be
\Theta_{02}\colon\quad\left\{
\begin{array}{rcl}
 \upsilon_0 &\longmapsto& \hf \left(f_3+f_5+f_6-f_9+f_{15}-f_{19}-f_{23}-f_{24}\right),\\[5pt]
 \upsilon &\longmapsto&  \hf \left(f_3+f_5+f_6-f_9-f_{15}-f_{19}-f_{23}-f_{24}\right).
\end{array}\right.
\ee
This completes the construction of the overarching map $\Theta_{02}$ for the square and the triangular
Kummer surfaces.
Again, the overarching map $\Theta_{02}$ 
leads to an overarching symmetry group, whose action is encoded in the same Niemeier lattice 
$N(-1)$ through the generators \eqref{genericc24} of the translational symmetry group $G_t$ common to all 
Kummer surfaces, in addition to
the generators \eqref{alphapermutations} {\em and} \eqref{betapermutations}. 
The resulting group is a copy of 
$(\mathbb{Z}_2)^4\rtimes \DDD \subset M_{24}$, where $\DDD$ denotes the binary dihedral
group of order $12$, as before. 

%%%%%%%%%%%%%%%%%%%%%%%%%%%%%%%%%%%%%%%%
\subsection{Overarching the tetrahedral and triangular Kummer K3s}
%%%%%%%%%%%%%%%%%%%%%%%%%%%%%%%%%%%%%%%%
The construction of an overarching map $\Theta_{12}$ for $X_1$ and $X_2$ 
requires a root $f_{n_0}$ with $n_0\in\OOO_9$ which is 
 invariant under $G_1$ and $G_2$,
whose generators are given in \req{betapermutations} and \req{gammapermutations}.
The only label in $\OOO_9$ which is invariant under both these groups is $n_0=6$.

The generators of the rank $3$ lattice $\wh K_{G_1}$ are given  in \eqref{Iperp01}, and those of the lattice 
$\wh K_{G_2}$ by \eqref{Iperp02}.
From \cite[(4.3)]{tawe11} we read that $n_0=6\not\in Q_{ij}$ implies
\be \label{Qij6}
\begin{array}{rclrclrcl}
Q_{12}&=&\{5, 9, 23, 24\}, 
&\quad Q_{13}&=&\{3, 5, 9, 19\},
&\quad Q_{14}&=&\{3, 9, 15, 24\}, \\[5pt]
Q_{34}&=&\{3, 5, 23, 24\}, 
 &\quad  Q_{24}&=&\{15, 19, 23, 24\}, 
&\quad  Q_{23}&=&\{3, 9, 15, 23\}. 
\end{array}
\ee
Hence the map $\qu I$ described in Section\,\ref{sec:overarching} is
$$
\begin{array}{rclrclrcl}
\qu I(I_{1,1}^{\perp}) &=& \{ 15, 19\}, 
&\quad \qu I(I_{2,1}^{\perp}) &=& \{ 9, 15\},
&\quad \qu I( I_{3,1}^{\perp}) &=& \{ 15, 24\},\\[5pt]
\qu I( I_{1,2}^{\perp}) &=& \{ 3, 9\},
&\quad  \qu I( I_{2,2}^{\perp}) &=& \{ 5, 24\},
 &\quad \qu I( I_{3,2}^{\perp}) &=& \{ 23, 24\}.
\end{array}
$$
We now need to choose the images in $N$ of the generators   $I^\perp_{i,1},\, i\in\{1,2,3\}$ 
and  $I^\perp_{i,2},\, i\in\{1,2,3\}$ under $\Theta_{12}$ such that $\Theta_{12}$ restricts to an 
isometry on the lattices 
$\wh K_{G_1}$ and $\wh K_{G_2}$. 
Given the quadratic form \eqref{quadform0} for $\wh K_{G_1}$ and \eqref{quadform2}
for $\wh K_{G_2}$, the following gives linearly independent candidates for $\Theta_{12}( I_{i_k, k}^\perp )\in N$:
\begin{equation} \label{theta12}
\Theta_{12}\colon\quad\left\{
\begin{array}{rcrrcrrcr}
 I_{1,1}^{\perp} &\longmapsto& f_{19}-f_{15},
&\quad  I_{2,1}^{\perp}&\longmapsto& f_{9}-f_{15},
&\quad I_{3,1}^{\perp} &\longmapsto& f_{24}-f_{15},\\[5pt]
I_{1,2}^{\perp}&\longmapsto&   f_{3}-f_{9},
 &\quad I_{2,2}^{\perp}&\longmapsto& f_5-f_{24},
 &\quad I_{3,2}^{\perp}&\longmapsto& f_{24}-f_{23}.
\end{array}
\right.
\end{equation}
Equivalently,
\begin{equation}\label{theta12p}
\Theta_{12}\colon\quad\left\{
\begin{array}{rcrcl}
 \pi_\ast \lambda_{12}
&\longmapsto&  2q_{12}+2f_3-2f_{15}&=& -f_5-f_9+f_{23}+f_{24}+2f_3-2f_{15},\\[5pt]
 \pi_\ast \lambda_{34}
&\longmapsto& 2q_{34}-2f_{15}&=&f_3-f_5+f_{23}+f_{24}-2f_{15},\\[5pt]
 \pi_\ast \lambda_{13} 
&\longmapsto& 2q_{13}+2f_{15}-2f_{23}&=&-f_3+f_5+f_9-f_{19}+2f_{15}-2f_{23},\\[5pt]
\pi_\ast \lambda_{24}
&\longmapsto& 2q_{24}&=&-f_{15}+f_{19}+f_{23}-f_{24},\\[5pt]
 \pi_\ast \lambda_{14}
&\longmapsto& 2q_{14}&=&f_{3}-f_{9}-f_{15}+f_{24},\\[5pt]
 \pi_\ast \lambda_{23}
&\longmapsto& 2q_{23}&=&f_{3}-f_{9}-f_{15}+f_{23}.
\end{array}
\right.
\end{equation}
%
%where the $q_{ij}$ are defined analogously to \cite[(2.15)]{tawe11}.
On the Kummer lattice $\Pi$, we set $\Theta_{12}(E_{\vec a})=f_{I^{-1}(\vec a)}$.
Finally, a consistent choice for the images of $\upsilon,\, \upsilon_0$ is
\be
\Theta_{12}\colon\quad\left\{
\begin{array}{rcl}
 \upsilon_0 &\longmapsto& \hf \left(f_3+f_5+f_6-f_9-f_{15}-f_{19}+f_{23}-f_{24}\right),\\[5pt]
 \upsilon &\longmapsto&  \hf \left(f_3+f_5-f_6-f_9-f_{15}-f_{19}+f_{23}-f_{24}\right).
\end{array}
\right.
\ee
This completes the construction of the overarching map $\Theta_{12}$ which is compatible with the 
symmetry groups of the tetrahedral ($G_1$) and triangular $(G_2)$ Kummer surfaces. 
Hence there is an overarching symmetry group
for the tetrahedral and the triangular Kummer surfaces, whose action is encoded in the same 
Niemeier lattice $N(-1)$ through
the generators \eqref{genericc24} of the translational symmetry group $G_t$ common to all Kummer surfaces, 
in addition to
the generators \eqref{gammapermutations} {\em and} \eqref{betapermutations}. The group 
thus generated is  a copy of 
$(\mathbb{Z}_2)^4\rtimes A_7 \subset M_{24}$. 
%
%%%%%%%%%%%%%%%%%%%%%%%%%%%%%%%%%%%%%%%%%%%%%%%%%%%%%%%%%%%%%%%
\section{ Overarching the moduli space  of Kummer K3s by  $\mathbf{(\Z_2)^4\rtimes A_8}$}
\label{sec:24A8}
%%%%%%%%%%%%%%%%%%%%%%%%%%%%%%%%%%%%%%%%%%%%%%%%%%%%%%%%%%%%%%
In this section we argue that our surfing procedure allows us to surf between any two points
of the moduli space of Kummer K3s. More precisely, for
any two Kummer surfaces with induced dual K\"ahler class, we can find representatives in the
smooth universal cover of the moduli space of hyperk\"ahler structures, such that an overarching
map between the two representatives exists. This allows us to combine
all symmetry groups of such Kummer surfaces to a larger, overarching group.\\ 

To see this, let us first consider an arbitrary Kummer surface $X_{\wt\Lambda,\omega_0}$ with 
induced dual K\"ahler class,
and let $\wt G$ denote its symmetry group. According to our discussion in 
Section\,\ref{sec:quaternions}, $\wt G=(\Z_2)^4\rtimes (\wt G_T^\prime/\Z_2)$, where
$\wt G_T^\prime\subset SU(2)$ is the linear automorphism group of the lattice $\wt\Lambda$. Moreover, 
$\wt G_T^\prime$ is a subgroup of one of the three
maximal linear automorphism groups of complex tori, the binary tetrahedral group $\TTT$ or one of the
dihedral groups $\DDD,\,\OOO$ of order $12$ and $8$. 

Let $G_T^\prime=\OOO, \TTT$ or $\DDD$,  
such that $\wt G_T^\prime\subset G_T^\prime$, and let $\Lambda=\Lambda_0,\, \Lambda_1$ 
or $\Lambda_2$ denote the corresponding
choice of lattice from Section\,\ref{sec:quaternions}
which has $G_T^\prime$ as its linear automorphism group. Fujiki's classification \cite{fu88}
implies that we can choose $G_T^\prime$ and $\Lambda$ in such a way that
there is a smooth deformation of $\Lambda$ into $\wt\Lambda$, {call it} $\Lambda^t$
with $t\in[0,1]$ and $\Lambda^0=\Lambda,\,\Lambda^1=\wt\Lambda$, such that the linear automorphism
group of each $\Lambda^t$ with $t\neq0$ is $\wt G_T^\prime$. The quaternionic language 
introduced in Section\,\ref{sec:quaternions} is particularly useful to check this.
For example, if $\wt G_T^\prime=\Z_4$,
then by Fujiki's results we can choose coordinates such that
the action of this group on $\C^2$ is generated by our symmetry $\alpha_1$ of \req{generatorssquare},
and we can choose $G_T^\prime=\OOO$ with $\Lambda=\Lambda_0$ the lattice of the square torus. One finds generators
$\vec\lambda_1^t,\,\ldots,\,\vec\lambda_4^t$ for the lattices $\Lambda^t$ as desired such that 
$\vec\lambda_2^t=\alpha_1(\vec\lambda_1^t)$ and $\vec\lambda_3^t=\alpha_1(\vec\lambda_4^t)$
for every $t\in [0,1]$. 

This deformation argument implies that by use of our fixed
marking, the invariant sublattices
of the integral torus homology, $L^{\wt G_T^\prime}=H_\ast(T,\Z)^{\wt G_T^\prime}$ and 
$L^{G_T^\prime}=H_\ast(T,\Z)^{G_T^\prime}$, obey $L^{G_T^\prime}\subset L^{\wt G_T^\prime}$.
Hence for the symmetry group $G$ of $X_{\Lambda,\omega_0}$ and
for the lattices that yield our Niemeier markings we have $M_{\wt G}\subset M_G$, see
Def.~\ref{niemeiermarking} and the discussion preceding it. 
From this it follows that one can find a representative of $X_{\wt\Lambda,\omega_0}$ in the smooth universal cover of
the moduli space of hyperk\"ahler structures such that every Niemeier marking 
$\iota_G\colon M_G\hookrightarrow N(-1)$ of the maximally symmetric Kummer surface $X_{\Lambda,\omega_0}$
restricts to a Niemeier marking $\iota_{\wt G}:=\iota_G{}_{\mid M_{\wt G}}$ of the Kummer 
surface $X_{\wt\Lambda,\omega_0}$. Hence any overarching map $\Theta$ for the maximally symmetric
Kummer surface $X_{\Lambda,\omega_0}$ and any other Kummer K3 $X$ also allows to surf from $X_{\wt\Lambda,\omega_0}$ to $X$.\\

Now consider two distinct Kummer surfaces $\wt X_A$ and $\wt X_B$ with their induced 
dual K\"ahler classes. By the above, we can choose maximally symmetric Kummer surfaces
$X_A$ and $X_B$ from the square, the tetrahedral  and the triangular Kummer surfaces, such
that the following holds: there are representatives of $\wt X_A$ and $\wt X_B$ in the smooth
universal cover of the moduli space of hyperk\"ahler structures such that any Niemeier marking
of $X_A$ restricts to a Niemeier marking of $\wt X_A$, and analogously for $X_B$ and $\wt X_B$.
Then by the above, the overarching map $\Theta_{AB}$ for $X_A$ and $X_B$ which was 
constructed\footnote{If $X_A=X_B$, then there is nothing left to be shown.}
in Section\,\ref{sec:constructionmaps} also overarches $\wt X_A$ and $\wt X_B$. In other words,
we can surf from $\wt X_A$ to $\wt X_B$.\\

We conclude that by means of our overarching maps we can surf the entire moduli space
of hyperk\"ahler structures
of Kummer surfaces.
In particular, we can combine the actions of all symmetry groups of Kummer surfaces with
induced dual K\"ahler class by means of their action on the Niemeier lattice $N$. Recall from 
Section\,\ref{sec:overarching} that by construction, every overarching map $\Theta_{AB}$
between Kummer surfaces $X_A$ and $X_B$ with symmetry groups $G_A$ and $G_B$
assigns a fixed root $\Theta_{AB}(\upsilon_0-\upsilon) = f_{n_0}\in N(-1)$ to the root
$\upsilon_0-\upsilon\in H_\ast(X,\Z)$, where $n_0\in\OOO_9$ is a label in our reference
octad from the Golay code\footnote{This fixed label $n_0$ is responsible for the
fact that each $G_k$ is a subgroup of $M_{23}$, as we emphasized in \cite{tawe11}.}. 
This root $f_{n_0}$ is fixed under the induced actions of both $G_A$ and $G_B$.
For the overarching group $G_{AB}$  obtained from $G_A$ and $G_B$, which by construction is a subgroup of the
stabilizer group $(\Z_2)^4\rtimes A_8$ of the octad $\OOO_9$, 
 this implies
that $G_{AB}$ additionally  fixes one label $n_0\in\OOO_9$. 
Hence $G_{AB}$ is a subgroup of  $(\Z_2)^4\rtimes A_7$, 
the group which we call
the \textsc{overarching symmetry group of Kummer K3s}. 
In Section\,\ref{sec:constructionmaps} we have seen that for two pairs of distinct Kummer 
surfaces with maximal symmetry, the overarching group yields $G_{AB}=(\Z_2)^4\rtimes A_7$. 
The third pair has overarching group
$(\Z_2)^4\rtimes \DDD$.
Moreover, in each case there exists precisely one label in $\OOO_9$ which is fixed
by both $G_A$ and $G_B$. This label, however, is different for each of the three pairs
of Kummer K3s with maximal symmetry. It follows that the combined symmetry group
for all Kummer K3s with induced dual K\"ahler class is $(\Z_2)^4\rtimes A_8$. 
%
%
%
%%%%%%%%%%%%%%%%%%%%%%%%%%%%%%%%%%%%%%%%%%%%%%%%%%%%%%%%%%%%%%%
\section{Interpretation and outlook}
\label{sec:interpretation}
%%%%%%%%%%%%%%%%%%%%%%%%%%%%%%%%%%%%%%%%%%%%%%%%%%%%%%%%%%%%%%
Let us now explain how our construction fits into 
the quest for the expected representation
of $M_{24}$ on a vertex algebra which governs the elliptic genus of K3. 
As mentioned in the Introduction, the elliptic genus arises from
the contribution to the partition function of any superconformal field theory on K3
which counts states in the Ramond-Ramond sector with signs according to 
fermion numbers. This part of the partition function is modular invariant on its own, inducing the
well-known modularity properties of the elliptic genus. The very construction of the elliptic genus, in addition, amounts to a
projection onto those states which are Ramond ground states
on the antiholomorphic side.
The usual rules for fermion
numbers imply that the OPE between any two fields in the Ramond sector yields contributions
from the Neveu-Schwarz sector only. Hence the expected vertex algebra can certainly not
arise in the Ramond-Ramond sector.
Of course we can spectral flow the relevant fields into the Neveu-Schwarz sector, where
(prior to all projections)
they indeed form a closed vertex algebra\footnote{Here and in the following, we loosely refer
to the space of fields which create states in the Neveu-Schwarz sector, equipped with the OPE, 
as a ``vertex algebra'', which
however is \emph{not} a holomorphic VOA.} $\wh\XXX$. Note that the choice of a spectral flow requires
the choice of a holomorphic and an antiholomorphic $U(1)$-current within the superconformal
algebra of our SCFT. For definiteness, we use the spectral flow which maps
Ramond-Ramond ground states to (chiral, chiral) states.\\

The resulting vertex algebra $\wh\XXX$ certainly governs the elliptic genus. Its space of states contains the
states underlying the
well-known (chiral, chiral) algebra $\XXX$ of Lerche-Vafa-Warner \cite{lvw89}, which accounts for
the contributions to the lowest
order terms of the elliptic genus.  In Appendix\,\ref{sec:transcft} we
describe the (chiral, chiral) algebra $\XXX$ (see \req{newbasis}) more concretely in the context 
relevant to this work, namely in $\Z_2$-orbifold conformal field theories 
$\CCC=\TTT/\Z_2$ on K3, where $\TTT$ denotes the underlying toroidal theory. 
As expanded upon in Appendix\,\ref{sec:transcft}, the 
very truncation to the (chiral, chiral) algebra $\XXX$ makes 
$\XXX$ completely independent of all 
moduli. In principle, this is a desired effect when aiming at constructing a vertex algebra which governs
the elliptic genus, since the latter is independent of all moduli. 
However, from the
action of a linear map on $\XXX$ (generated by the fields in \req{newbasis}, independently of all moduli), 
it is not clear whether or not it is a symmetry, while
the Mathieu Moonshine phenomenon dictates that we consider symmetries of some underlying vertex algebra.

We shall come back to this `bottom up' discussion further down, 
but we first take a closer look at the `top-down' approach, and
consider the action of symmetries of ${\CCC}$ on the 
(chiral, chiral) algebra $\XXX$ generated by \req{newbasis}.
We impose a number of rather severe assumptions on such symmetries,
in order to ensure that they descend to symmetries of a candidate
vertex algebra that governs the elliptic genus.
As mentioned in the Introduction, this graded vertex algebra at 
leading order  is the (chiral,chiral) algebra $\XXX$.
Following \cite{lvw89} we identify $\XXX$ with the cohomology of a K3 surface $X$.
Associated to every Calabi-Yau manifold $Y$, there is the chiral de Rham complex \cite{msv98}
which furnishes a sheaf of vertex algebras governing
the elliptic genus of $Y$ and containing  the {usual de Rham complex} of $Y$  at leading order {\cite{boli00,bo01}}.
We thus find it natural\footnote{By {\cite{boli00,frsz07}}, the CFT orbifold procedure descends to the
chiral de Rham complex; this should be the source for the behavior of the twining genera in
Mathieu Moonshine, at least for those symmetries that are induced from geometric ones.}
 to restrict our attention to symmetries of $\CCC$ that descend to
 the chiral de Rham complex of $X$.
  To this end, 
we assume that  our SCFT $\CCC$ 
comes with a choice of generators of the $N=(4,4)$ superconformal
algebra, which in particular fixes the $U(1)$-currents and a preferred $N=(2,2)$ subalgebra.
As mentioned above, this is already necessary when we choose the spectral flow to
$\XXX$. Recall that the choice of $U(1)$-currents amounts to the choice of a complex structure in
any geometric interpretation of $\CCC$ \cite{asmo94}. 
We furthermore use the notion advertised by \cite{gprv12}, which requires symmetries
to fix the superconformal algebra of $\CCC$ pointwise.\footnote{This, for instance,
excludes equivalences of SCFTs
induced by mirror symmetry, which acts as an outer automorphism on the superconformal
algebra.} To identify $\XXX$ with the  cohomology of a K3 surface $X$, we need to perform a large volume
limit \cite{wi82,lvw89}. {More generally}, 
according to \cite{ka05}, the space of states singled out by the elliptic
genus is mapped to the appropriate cohomology of the
chiral de Rham complex of $X$ only in the large volume limit. In order to perform such a 
large volume limit, we need to choose
a geometric interpretation of $\CCC$. 

Summarising, in view of constructing a vertex algebra from the fields in  $\CCC$, 
such that   $\XXX$ governs the leading 
order terms of the elliptic genus,  we  restrict our attention to symmetries
that are induced from geometric symmetries. This justifies why so far, in our work,
we have searched for explanations of Mathieu Moonshine phenomena within the
context of geometric symmetries only.

As a further potential justification for this restriction recall the notion of 
 `exceptional' symmetry
groups of sigma models on K3, that is, symmetry groups {of such SCFTs} which are not realizable as subgroups of $M_{24}$,
obtained from the classification in
 \cite{ghv10b}. According to \cite{gavo12}, in many cases `exceptional' symmetry is linked 
to certain quantum
symmetries which as we shall argue cannot be
 induced from any classical geometric symmetries. Indeed, these symmetries in \cite{gavo12} are
 characterized by the property that they generate a group $G$, such that orbifolding the K3 model by $G$
 yields a toroidal SCFT. We remark that there is no geometric counterpart of such an orbifold construction, which
 would have to yield a complex four-torus as an orbifold of a K3 surface. Indeed, the odd cohomology
 of a complex four-torus cannot be restored by blowing up quotient singularities in an orbifold by
 a symplectic automorphism group of a K3 surface. However, this is only a potential justification 
 for our restriction to geometric symmetries since,
 according to \cite{gavo12}, `exceptional' symmetry groups also occur in a few cases where to date
 it is not known whether or not such purely non-geometric quantum symmetries are responsible
 for the `exceptionality' of the symmetry group. Although the group $M_{24}$ itself contains elements 
 that can never act in terms of a geometric symmetry on K3, we are optimistic that every
 element of $M_{24}$ can be obtained as a composition of `geometric' symmetries.\\

We wish to emphasize that it is immediately clear that the (chiral, chiral) algebra
$\XXX$ cannot carry a representation of $M_{24}$. 
Indeed,  \req{fixedforms} is the basis of a four-dimensional
subspace of the $24$-dimensional space $\XXX$ which is invariant under all symmetries that are of
interest here, but by the known properties of representations of $M_{24}$, this group can only act trivially
on the remaining $20$-dimensional space. Hence a vertex algebra which governs the massless leading
order terms of the elliptic genus, and which at the same time carries the expected representation of $M_{24}$,
must be related to $\XXX$ by some nontrivial map. The Niemeier markings and the overarching maps which
were  constructed in \cite{tawe11} should be viewed as a first approach towards constructing
such a map. This claim is based on the observation that,
from a geometric viewpoint, the introduction of Niemeier markings is necessary
 to combine symmetry groups of Kummer surfaces into larger groups. Indeed,
it follows from Mukai's results  that for any finite group $\wh G$ of 
lattice automorphisms of $H_\ast(X,\Z)$ that is 
not a subgroup of one of the eleven maximal groups listed in \cite{mu88},
the lattice $L_{\wh G}:=(H_\ast(X,\Z)^{\wh G})^\perp\cap H_\ast(X,\Z)$
is indefinite and thus violates the signature requirements for symmetries of K3 surfaces. Therefore, we 
never expected $M_{24}$ to act on $H_\ast(X,\Z)$ either.
It would be interesting to see if the massive sector of the elliptic genus 
is also subject to a `no-go theorem' when working in the framework of $\Z_2$-orbifold 
CFTs on K3. A priori, the situation could be different, as the original Mathieu Moonshine
observation \cite{eot10} states that in the elliptic genus,
the multiplicities of {\em massive} characters of the $N=4$ superconformal
algebra yield dimensions of representations of $M_{24}$. 
In a forthcoming work \cite{tawe13}
we present evidence in favour of our expectation that the {\em massive} fields
which contribute to the elliptic genus are related to a representation of $M_{24}$
in a much more immediate fashion. \\ 

We now return to the `bottom-up' approach, and investigate more closely
 the action of symmetry groups on the vertex algebra 
 $\XXX$, to explain in terms of CFT data how our Niemeier markings 
 and overarching maps are relevant in the context of SCFTs on K3. 
To this end note that  the
entire group $\mbox{SL}(2,\C)$ acts naturally on the truncated vertex algebra $\C\otimes\XXX$ of 
\req{newbasis}, preserving
$U(1)$-charges. 
However, a given element of $\mbox{SL}(2,\C)$ may not have an extension to a symmetry
of the full SCFT $\CCC$. Whether or not this is the case cannot be determined from the action
on the fields listed in \req{newbasis}. 
Indeed, 
this depends on the moduli of $\CCC$, but
the vertex algebra  $\XXX$ has lost its  dependence on all moduli
due to the truncation, as described earlier.  However, as we explain in Appendix\,\ref{sec:transcft},
one may introduce the analog  $\XXX^\Z$ of the lattice
of integral homology in the vector space $\XXX$, and use its structure 
 to determine whether or not an element of $\mbox{SL}(2,\C)$
acts as a symmetry of $\CCC$. \\

By the above, we are only interested in symmetry groups $G$ 
that are induced by geometric symmetries,
and in line with our work so far, we restrict our attention to those that are induced\footnote{This
includes the symmetries induced by half lattice shifts in the underlying toroidal theory $\TTT$.} 
from the underlying toroidal CFT $\TTT$.
By definition, a symmetry of a SCFT must be compatible with all OPEs in that theory. 
In particular, the standardized OPE \req{standardope} must be preserved. 
Following the arguments presented in Appendix\,\ref{sec:transcft}, this implies that
each of
our symmetry groups $G$ acts as a group of lattice automorphisms on $\XXX^\Z$, such
that this lattice of fields in our SCFT contains a sublattice $\XXX^\Z_G$ which
bears all relevant information about the $G$-action on our
SCFT. This lattice  can be identified with the lattice $M_G$ which is central to our construction,
in that our Niemeier markings isometrically replicate it as a sublattice of the Niemeier lattice $N(-1)$.
This allows a more elegant description of  $G$ as a subgroup of $M_{24}$, and it enables
us to combine the symmetry groups from distinct K3 theories to a larger, overarching group. 
In other words, our Niemeier marking describes precisely the action of geometric symmetry groups
on the vertex algebra  which governs the elliptic genus to leading order
terms.
This justifies the relevance of our construction in the context
of our quest to unravel some of the mysteries of the Mathieu Moonshine phenomenon.
\\

The picture that we offer  here shows how the beautiful interplay between
geometry and conformal field theory may yield some keys to the Mathieu Moonshine Mysteries.
Such an interplay is  expected. On the one hand, the elliptic genus is a purely geometric quantity. On the other hand,
this quantity also appears in the context of SCFTs on K3, 
where its decomposition into $N=4$ characters is natural. 
Notably, it is only after decomposing the elliptic genus into $N=4$ characters that one  
observes the
Mathieu Moonshine phenomenon \cite{eot10}.

We expect that order by order, the elliptic genus dictates the construction of 
representations of $M_{24}$ on appropriately
truncated vertex algebras arising from SCFTs on K3. In other words, the very representations
of $M_{24}$ that are observed in the elliptic genus are intrinsic to these SCFTs. 
The reason why the emerging group is $M_{24}$ is still unclear, but 
we expect it to be
rooted in the structure of these SCFTs, where
geometry dictates the symmetries which can act on these representations. 
By symmetry-surfing the moduli space of SCFTs on K3, 
we expect that the natural representations of geometric symmetry
groups on these vertex algebras combine to the action of $M_{24}$.\\

Our construction of overarching maps in \cite{tawe11} should be viewed as a very first step towards
finding such vertex algebras for the leading order
terms of the elliptic genus.
In the present work, we show that our overarching maps indeed allow us to combine all relevant 
symmetry groups, 
as long as we restrict to $\Z_2$-orbifold conformal
field theories on K3 and their geometric interpretations on Kummer K3s,
and to symmetries that are induced geometrically from the underlying toroidal theories.
Indeed, since 
one can easily associate a vertex algebra to the Niemeier lattice $N$, one could claim that we have solved the
problem of constructing a vertex algebra that furnishes the expected symmetries.
However, of course we pay dearly since this vertex algebra does
not govern the leading order terms of the elliptic genus in any obvious way. Still
our approach paves the way to defining the desired vertex algebra.
As we have explained above, we expect
 vertex algebras associated  with all  remaining orders of the elliptic
genus to relate directly to the respective representations of $M_{24}$, and we present evidence in favour of this
expectation in \cite{tawe13}.

%%%%%%%%%%%%%%%%%%%%%%%%%%%%%%%%%%%%%%%%%%%%%%%%%%%%%%%%%%%%%%%
\appendix
\section{Transition to superconformal field theory}\label{sec:transcft}
%%%%%%%%%%%%%%%%%%%%%%%%%%%%%%%%%%%%%%%%%%%%%%%%%%%%%%%%%%%%%%
Throughout our work, we use homological data to describe geometric symmetries of K3 surfaces.
This is natural, since the techniques are well-established in algebraic geometry, but also since
the well-known properties of (chiral, chiral) algebras \cite{wi82,lvw89} recover (co)homological 
data from sigma model interpretations of SCFTs. 
This is particularly straightforward for the $\Z_2$-orbifold conformal field theories which are
relevant to our investigations.
Since our work
is motivated by Mathieu Moonshine \cite{eot10}, which is rooted in conformal field theory,
and since the role of the {\em integral} (co)homology in
(chiral, chiral) algebras seems not so well established, we gather in this appendix the tools needed
to make a smooth transition 
to superconformal field theory.\\

We first need to fix some notations.
Every toroidal conformal field theory $\TTT$
possesses two free Dirac fermions on the holomorphic side,
which we denote by $\chi^1_+(z),\, \chi^2_+(z)$. The complex conjugate fields are denoted 
$\chi^1_-(z),\, \chi^2_-(z)$, such that
$$
\chi^i_+(z)\chi^j_-(w)\sim{\delta_{ij}\over z-w},\qquad i,j \in\{1,2\},
$$
while the antiholomorphic counterparts are denoted $\qu\chi^1_\pm(\qu z),\, 
\qu\chi^2_\pm(\qu z)$.
The corresponding holomorphic - antiholomorphic combinations are more
appropriate for our purposes,
$$
\xi_1:={1\over2}\left(\chi^1_++\qu\chi^1_+\right),\quad
\xi_2:={1\over2i}\left(\chi^1_+-\qu\chi^1_+\right),\quad
\xi_3:={1\over2}\left(\chi^2_++\qu\chi^2_+\right),\quad
\xi_4:={1\over2i}\left(\chi^2_+-\qu\chi^2_+\right).
$$
Moreover, in every $\Z_2$-orbifold conformal field theory $\CCC=\TTT/\Z_2$ on K3,
there is a $16$-dimensional space of twisted ground
states, generated by fields $T_{\vec a}$  in the Ramond-Ramond sector, where the label
$\vec a\in\F_2^4$ refers to the fixed point $\vec F_{\vec a}$ as in \req{labels}
at which the respective field is localized. For ease of notation
we denote by $\wt T_{\vec a},\, {\vec a}\in\F_2^4$, the 
(chiral, chiral) fields which the $T_{\vec a}$ flow to under our choice of 
spectral flow. 
Then the following is a list of  $24$ fields which generate
the (chiral, chiral) algebra in every  theory $\CCC=\TTT/\Z_2$ on K3:
\be\label{newbasis}
\xi_1\xi_2\xi_3\xi_4,\quad \xi_i\xi_j \;(1\leq i<j\leq 4),\quad \id;\quad \wt T_{\vec a}\; ({\vec a}\in\F_2^4),
\ee
where $\id$ denotes the vacuum field,
and where we may restrict our attention to the real vector space $\XXX$
generated by these $24$ fields.
After truncation of the OPE to chiral primaries \cite{lvw89}, the fields listed in (\ref{newbasis})
form a closed vertex algebra  $\XXX$ over $\R$. Note that this very truncation makes the vertex algebra 
completely independent of all 
moduli.  

We remark that the real and imaginary parts\footnote{Here and in the following, for a 
field $\eta\in\C\otimes\XXX$ with $\eta=\eta_1+i\eta_2$ and $\eta_1,\,\eta_2\in\XXX$, we call
$\eta_1,\,\eta_2$ the real and the imaginary part of $\eta$.} of the four fields 
with $U(1)$-charges $(2,2),\,(2,0)$, $(0,2),\,(0,0)$ in \req{newbasis}, 
\be\label{fixedforms}
\xi_1\xi_2\xi_3\xi_4,\; \xi_1\xi_3-\xi_2\xi_4,\;  \xi_1\xi_4+\xi_2\xi_3,\;\id,
\ee
remain invariant under every symmetry of $\CCC$. These fields are naturally identified with the cycles 
$\pi_\ast\upsilon^T$,\, $\Omega_1,\,\Omega_2$, $\pi_\ast \upsilon_0^T\in\pi_\ast H_\ast(T,\R)$ on K3, 
with $\Omega_1,\,\Omega_2$ as in \req{invariant} and $\upsilon^T,\,\upsilon^T_0$ generators
of $H_4(T,\Z)$ and $H_0(T,\Z)$ such that $\langle\upsilon^T,\upsilon^T_0\rangle=1$.
The invariance of $\Omega_1,\,\Omega_2$
under symmetries means that in a given geometric interpretation,  
one restricts attention to symplectic automorphisms (see \cite{tawe11} for further details).
In the  description of the moduli space of  SCFTs on a K3 surface $X$ of \cite{asmo94}, 
our SCFT $\CCC$ is specified by the relative position of a positive definite fourplane in $H^*(X,\R)$
with respect to $H^*(X,\Z)$. The two-forms $\Omega_1,\,\Omega_2$ generate a two-dimensional
subspace of that fourplane, while the choice of $\upsilon^T$
and $\upsilon_0^T$   amounts to the choice of
a geometric interpretation of the toroidal theory $\TTT$ which induces a natural
geometric interpretation of its $\Z_2$-orbifold $\CCC$ (see \cite{nawe00,we00}). 
Here, the four fields in \req{fixedforms} are the real and imaginary parts
of the images of the four charged Ramond-Ramond ground
states under our choice of spectral flow. These four Ramond-Ramond ground states also furnish
a fourplane that can be used to describe the moduli space of superconformal field theories on K3
\cite{nawe00}.
Note however that the fourplane of \cite{asmo94} is  not the one generated by the 
four vectors in \req{fixedforms}.

The vector space $\XXX$ can be identified with the real K3 homology
$H_\ast(X,\R)$, where the
$\xi_i\xi_j$ with $1\leq i<j\leq 4$ are mapped to our generators $e_i\vee e_j$ of $\pi_\ast H_2(T,\R)$,
and the $\wt T_{\vec a}$ are in $1\colon1$ correspondence with the cycles $E_{\vec a}$
that arise from the minimal resolution of $T/\Z_2$ (see \cite{nawe01} for the subtleties 
in this identification, due to the B-field that is induced by orbifolding). 

One may, in addition, introduce the analog  of the lattice
of integral homology for the vector space $\XXX$, thereby recovering the dependence on the moduli. 
To appreciate this, note that before truncation 
the OPE between twist fields $T_{\vec b}$ and $T_{\vec b^\prime}$
with $\vec b,\,\vec b^\prime\in\F_2^4$  yields, to leading order, a primary field 
$W_{\vec b-\vec b^\prime}(z,\qu z)$ which does depend on the moduli.  This is best measured by
means of the OPE between the free bosonic superpartners of the Dirac fermions $\xi_1,\ldots,\,\xi_4$
and $W_{\vec b-\vec b^\prime}(z,\qu z)$. For convenience of notation, we introduce real, holomorphic
$U(1)$-currents $j_1(z),\ldots, j_4(z)$, which arise as the superpartners of the real and the imaginary parts
of $2\chi^1_+(z),\,2\chi^2_+(z)$, respectively, and note that the relevant OPE then has the form 
$$
j_k(z) W_{\vec a}(w,\qu w)\sim { W_{\vec a}(w,\qu w)\over z-w} \sum_{l=1}^4 a_l \lambda^l_k
\quad\mbox{ for }\vec a=( a_1,\ldots,a_4)\in\F_2^4.
$$
Here, $\lambda^l_1,\ldots,\lambda^l_4$ are the Euclidean coordinates of generators
$\vec\lambda_1,\ldots,\vec\lambda_4$ of a rank $4$ lattice $\Lambda\subset\R^4$,
if the underlying toroidal SCFT $\TTT$ has a geometric interpretation on the torus 
$T=\R^4/\Lambda$,
where we identify $\R^4$ with $\C^2$ as usual. 
We observe that in the truncation procedure yielding the (chiral, chiral) algebra  
$\XXX$ of \req{newbasis},  the moduli-dependent fields $W_{\vec a}(z,\qu z)$ are projected to zero, 
and therefore the dependence on the moduli disappears from  $\XXX$. 
However, one may introduce new fields
\be\label{changeofbasis}
J_k(z):=\sum_{l=1}^4 \mu^l_k j_l(z)
\qquad\mbox{ for } k\in\{1,\ldots,4\},
\ee
where $\mu^l_1,\ldots,\mu^l_4$ are the Euclidean coordinates of
the basis $\vec\mu_1,\ldots,\vec\mu_4$  dual\footnote{Here, we identify $\R^4\cong(\R^4)^\ast$ by means
of the standard Euclidean scalar product.} to $\vec\lambda_1,\ldots,\vec\lambda_4$, 
such that the OPEs with the fields $W_{\vec a}(w,\qu w)$ 
take the standardized ``integral''
form 
\be\label{standardope}
J_k(z) W_{\vec a}(w,\qu w)\sim {a_k\over z-w} W_{\vec a}(w,\qu w),\qquad k\in\{1,\ldots,4\}.
\ee
The fermionic superpartners $\wt\Psi_1(z),\ldots,\wt\Psi_{4}(z)$ of the new fields
$J_1(z),\ldots, J_4(z)$ and their antiholomorphic counterparts $\qu\Psi_1(\qu z),\ldots,\qu\Psi_{{4}}(\qu z)$
yield a lattice with generators 
$$
\wt\Psi_1\wt\Psi_2\wt\Psi_3\wt\Psi_4,\, \wt\Psi_k\wt\Psi_l\qu\Psi_m\qu\Psi_n, \,
\wt\Psi_k\qu\Psi_l, \ldots
$$
over $\Z$.
However, to determine a lattice which plays the role of the integral homology of the
Kummer surface $X$, one needs to recall that the identification\footnote{From \cite{lvw89}, we 
obtain an immediate identification with cohomology, which
however is equivalent to homology by Poincar\'e duality.} of $\XXX$ with $H^\ast(X,\R)$
rests on the correspondence $\chi_+^k\leftrightarrow dz_k,\; ,\qu\chi_+^k\leftrightarrow d\qu z_k$
for $k\in\{1,\,2\}$, with local holomorphic coordinates $z_1,\, z_2$ on $X$. This correspondence
holds exactly on flat manifolds and in a large radius limit \cite{wi82,lvw89}. Hence at large
radii, the {\em real}\footnote{For open strings, one 
can view $\chi^k_+$ and its antiholomorphic partner $\qu\chi^k_+$ as complex conjugates, 
where the left and right modes combine into standing waves. In this language, we are simply
reviewing the emergence of charge lattices for D-branes.} fermionic fields 
$\wt\Psi_k$ are identified with the $\qu\Psi_k$, and thus with 
$$
\Psi_k :=  \sum_{l=1}^4 \mu^l_k \xi_l
\qquad\mbox{ for } k\in\{1,\ldots,4\}.
$$
This leaves us with the lattice $\YYY^\Z$ generated over $\Z$ by
$$
\Psi_1\Psi_2\Psi_3\Psi_4,\quad \Psi_i\Psi_j, \;(1\leq i<j\leq 4),\quad \id,
$$
which is the analog of the lattice 
$\pi_\ast H_\ast(T,\Z)\subset H_\ast(X,\R)$. Using the twist fields $\wt T_{\vec a}, \vec a\in\F_2^4$,
as additional generators that correspond to the  vectors $E_{\vec a}$ in the Kummer lattice, and then
performing the usual gluing procedure, one obtains a lattice $\XXX^\Z$ which can be 
identified
with $H_\ast(X,\Z)\subset H_\ast(X,\R)$.
In particular, the relative position of $\XXX^\Z$ with respect to
the basis \req{newbasis} of $\XXX$ determines the respective point in the moduli space. For the SCFT 
associated with the square
Kummer surface\footnote{with vanishing B-field on the underlying toroidal theory}, 
we can choose the eight fields $\xi_1\xi_2\xi_3\xi_4,\; \xi_i\xi_j \;(1\leq i<j\leq 4),\; \id$
as generators of the lattice $\YYY^\Z$.\\

Now note that each symmetry of a Kummer surface $X_{\Lambda,\omega_0}$
as studied in our work induces a symmetry of a SCFT $\CCC=\TTT/\Z_2$,
with $\TTT$ a toroidal theory\footnote{This leaves  a choice of the B-field $B_T$ 
in the toroidal theory $\TTT$, which must be
invariant under our symmetry; of course $B_T=0$ is always admissible.} associated with  the torus 
$\R^4/\Lambda$. By construction, our geometric symmetry groups $G$ 
 enjoy an induced action as groups of lattice automorphisms on the lattice $\XXX^\Z$.
By definition, the symmetries of a CFT are compatible with all OPEs, hence they must
 in particular leave the standardized OPEs \req{standardope} invariant.
Since our symmetries are induced by geometric symmetries
of the toroidal theory $\TTT$, they act linearly on
the $J_k(z)$ and they permute the fields $\pm W_{\vec a}(z,\qu z)$.
It follows that such symmetries act as lattice automorphisms on the lattice generated
by the $J_k(z)$. The same thus holds for the lattice generated by their 
superpartners $\Psi_k(z)$ and for the lattice $\YYY^\Z$ mentioned above. 
Since our symmetries also permute the twist fields 
$\pm\wt T_{\vec a}$ amongst each other in a manner compatible with gluing, 
altogether it follows that they must act as automorphisms of
the lattice $\XXX^\Z$.
By the above, the vector space $\XXX$ can
be identified with the K3 homology,  and $\XXX^\Z$ can be identified
with the integral homology. 
In particular, the lattice $\XXX^\Z$ possesses a sublattice
$\XXX^\Z_G$ which can be identified with the lattice $M_G$ that is so crucial to our construction,
see Def.~\ref{niemeiermarking}.
The action of $G$ on $\XXX^\Z_G$ bears all relevant information about the $G$-action on our
SCFT. Our construction hence realizes the very representation of $G$ on  $\XXX$ 
in terms of the action of a subgroup $G$ of $M_{24}$
on the Niemeier lattice $N$. 
In other words, our Niemeier marking describes precisely the action of the relevant symmetry groups
on the (chiral, chiral) algebra.

Certainly from the  description of the moduli space of SCFTs in terms of cohomological data
\cite{asmo94,nawe00}, we are lead to expect that the role of the (chiral, chiral) algebra $\XXX$
along with the  lattice
$\XXX^\Z$ in its underlying vector space  generalizes to arbitrary K3 theories.

\bibliographystyle{amsplain}
\def\polhk#1{\setbox0=\hbox{#1}{\ooalign{\hidewidth
  \lower1.5ex\hbox{`}\hidewidth\crcr\unhbox0}}} \def\cprime{$^\prime$}
\providecommand{\bysame}{\leavevmode\hbox to3em{\hrulefill}\thinspace}

\end{document}